\documentclass[11pt,draftcls,onecolumn]{IEEEtran}
\ifCLASSINFOpdf
\else
   \usepackage[dvips]{graphicx}
\fi
\usepackage{url}
\usepackage{graphicx}
\usepackage{amsmath}
\usepackage{comment}
\usepackage[utf8]{inputenc}
\usepackage[english]{babel}
\usepackage{color}
\usepackage{framed}

\usepackage{amsmath,amssymb,amsfonts}
\usepackage{graphicx}
\usepackage{upgreek}
\usepackage{textcomp}
\usepackage{url}

\usepackage{cite}
\usepackage{bbold}
\usepackage{ifpdf}
\usepackage{times}
\usepackage{layout}
\usepackage{float}
\usepackage{afterpage}
\usepackage{amsmath}
\usepackage{amstext}
\usepackage{amssymb,bm}
\usepackage{latexsym}
\usepackage{color}
\usepackage{flexisym}
\usepackage{kantlipsum}
\usepackage{graphicx}
\usepackage{amsmath}
\usepackage{amsthm}
\usepackage{graphicx}

\usepackage[caption=false,font=footnotesize]{subfig}

\usepackage{booktabs}
\usepackage{multicol}

\usepackage{enumitem}

\usepackage[switch]{lineno}
\usepackage[named]{algo}
\usepackage[noend]{algpseudocode}
\usepackage{algorithm}
\usepackage{amsmath}
\usepackage{amsthm}
\usepackage{dsfont}
\usepackage{thmtools}
\usepackage{amssymb}
\usepackage[dvipsnames]{xcolor}

\def\bb{\boldsymbol{b}}

\def\bp{\boldsymbol{p}}

\def\bI{\boldsymbol{I}}

\def\bGamma{\boldsymbol{\Gamma}}

\def\bDelta{\boldsymbol{\Delta}}
\def\bepsilon{\boldsymbol{\epsilon}}

\def\btheta{\boldsymbol{\theta}}

\def\bmu{\boldsymbol{\mu}}

\def\bSigma{\boldsymbol{\Sigma}}

\def\bOmega{\boldsymbol{\Omega}}

\def\mbb{\mathbf{b}}
\def\mbc{\mathbf{c}}

\def\mbe{\mathbf{e}}

\def\mbp{\mathbf{p}}

\def\mbr{\mathbf{r}}

\def\mbu{\mathbf{u}}

\def\mbx{\mathbf{x}}
\def\mby{\mathbf{y}}
\def\mbz{\mathbf{z}}

\def\mbA{\mathbf{A}}

\def\mbC{\mathbf{C}}

\def\mbR{\mathbf{R}}

\newtheorem{theorem}{Theorem}
\newtheorem{proposition}{Proposition}
\newtheorem{lemma}{Lemma}

\newtheorem{claim}{Claim}

\theoremstyle{definition}

\declaretheorem[style=definition,name=Remark,qed=$\blacksquare$]{remark}

\makeatletter
\newcommand*{\rom}[1]{\expandafter\@slowromancap\romannumeral #1@}
\makeatother

\newcommand{\Ae}[1]{\textcolor{blue}{#1}}

\begin{document}
% Reduce spacing above and below equations
% \setlength{\abovedisplayskip}{3pt}
% \setlength{\belowdisplayskip}{3pt}

\title{UNO: Unlimited Sampling Meets One-Bit Quantization}

\author{Arian Eamaz, \IEEEmembership{Graduate Student Member, IEEE}, Kumar Vijay Mishra, \IEEEmembership{Senior Member, IEEE}, Farhang Yeganegi, and Mojtaba Soltanalian, \IEEEmembership{Senior Member, IEEE}
\thanks{This work was supported in part by the National Science Foundation Grant CCF-1704401. The conference precursor to this work has been submitted for publication at the 2023 IEEE International Conference on Acoustics, Speech and Signal Processing (ICASSP)\\ (\emph{Corresponding author: Arian Eamaz})}
\thanks{A. Eamaz, F. Yeganegi, and M. Soltanalian are with the Department of Electrical and Computer Engineering, University of Illinois Chicago, Chicago, IL 60607, USA (e-mail: \emph{ \{aeamaz2, fyegan2, msol\}@uic.edu}).}
\thanks{K. V. Mishra is with the United States DEVCOM Army Research Laboratory, Adelphi, MD 20783 USA (e-mail: \emph{kvm@ieee.org}).}
%\thanks{\textcolor{red}{The conference precursor to this work has been submitted for publication at the 2023 IEEE International Conference on Acoustics, Speech and Signal Processing (ICASSP)} \textcolor{blue}{we need to disclose this and write in the journal paper the differences w.r.t. the conference paper. Please also consult M.S. about this.} } 
}

\markboth{Submitted to the IEEE TRANSACTIONS ON SIGNAL PROCESSING, 2022}
{Shell \MakeLowercase{\textit{et al.}}: Bare Demo of IEEEtran.cls for IEEE Journals}
\maketitle

\begin{abstract}
Recent results in one-bit sampling %has recently been widely adopted due to its 
provide a framework for a relatively low-cost, low-power sampling, at a high rate by employing time-varying sampling threshold sequences. Another recent development in sampling theory is unlimited sampling, which is a high-resolution technique that relies on \emph{self-reset ADCs} to yield an unlimited dynamic range. % in sampling. 
In this paper, we leverage the appealing attributes of the two aforementioned techniques to propose a novel %framework that enjoys the benefits of both one-bit and unlimited sampling approaches. To make this happens, we propose 
\emph{un}limited \emph{o}ne-bit (UNO) sampling approach. % which particularly addresses the  design of the sampling thresholds. Using this sampling approach, 
In this framework, the information on the distance between the input signal value and the threshold are stored and utilized to accurately reconstruct the one-bit sampled signal. We then utilize this information to accurately reconstruct the signal from its one-bit samples via
the randomized Kaczmarz algorithm (RKA); a strong linear feasibility solver that selects a random linear equation at each iteration. In the presence of noise, we employ the recent plug-and-play (PnP) priors technique with alternating direction method of multipliers (ADMM) to exploit integration of state-of-the-art regularizers in the reconstruction process. Numerical experiments with RKA and PnP-ADMM-based reconstruction illustrate the effectiveness of our proposed UNO, including its superior performance compared to the one-bit $\Sigma\Delta$ sampling.
\end{abstract}


\IEEEpeerreviewmaketitle

\section{Introduction}
\IEEEPARstart{S}{ampling} theory lies at the heart of all modern digital %information 
processing systems. The original sampling problem entails identifying a continuous function on Euclidean space from discrete data samples. It is addressed by the classical sampling theorem, commonly and variously, attributed to Cauchy \cite{cauchy1841memoire}, de La Vallée Poussin \cite{poussin1908sur}, Whittaker \cite{whittaker1915functions}, Ogura \cite{ogura1920certain}, Kotel\'{n}ikov \cite{kotelnikov1933on}, Raabe \cite{raabe1939untersuchungen}, Shannon \cite{shannon1949communication}, and/or Someya \cite{someya1949waveform}. A seminal result in this context, referred to as Whittaker-Kotel\'{n}ikov-Shannon (or, simply Shannon's) theorem, states that it is possible to fully recover a bandlimited function from values measured on a regular sampling grid as long as the bandlimitation is an interval with length not exceeding the density of the sampling grid. %A seminal result in the sampling literature is the Whittaker-Kotel\'{n}ikov-Shannon (or, simply Shannon's) theorem, which states that 
Restating this in signal processing terms, a lowpass bandlimited signal can be perfectly reconstructed from its discrete samples taken uniformly at a sampling frequency that is at least the Nyquist rate, i.e., twice the signal bandwidth. During the past few decades, several variants and extensions of this result have solidified the extensive role of sampling theory in science and engineering \cite{jerri1977shannon,hogan2012duration,pfander2015sampling}.

Shannon's theorem assumes the existence of samples that are of infinite precision and infinite dynamic range (DR). But, in practice, it is realized by the quantization of the signals through analog-to-digital converters (ADCs) that clip or saturate whenever the signal amplitude exceeds the maximum recordable ADC voltage, leading to a significant information loss. The effects of finite precision quantization are characterized in the form of rate distortion theory \cite{kailath1967application,berger2003rate}. However, investigations into finite DR or clipping effects are relatively recent \cite{olofsson2005deconvolution,adler2011audio,abel1991restoring,ting2013mitigation}. Substantial work has been done and is still ongoing to overcome this problem, and the literature is too large to summarize here; see, e.g., \cite{bhandari2020unlimited} and the references therein, for comparisons of various techniques. Overall, these approaches require declipping \cite{esqueda2016aliasing}, multiple ADCs \cite{gregers2001stacked}, and scaling techniques \cite{prasanna2021application}, which are expensive and cumbersome. Recently, some studies \cite{bhandari2020unlimited,bhandari2017unlimited,rudresh2018wavelet} have proposed the \emph{unlimited sampling} architecture to fully overcome this limitation by employing modular arithmetic. To perfectly reconstruct the signal of interest from modulo samples (up to a unknown constant), the unlimited sampling theory suggests the sampling rate to be slightly higher than the Nyquist rate and the norm estimate of the bandlimited signal be known.%must satisfy $T \leq \frac{1}{2\Omega_{\textrm{max}} e}$ where $\Omega_{\textrm{max}}$ denotes the maximum frequency of the input signal and $e$ is the Euler number. %\textcolor{red}{some details on what the condition is. See, e.g., my write up on page 3 before Theorem 1, where I mention that the norm estimate should be known} 
%for reconstruction of bandlimited signals from their modulo samples is proposed in \cite{bhandari2020unlimited} to guarantee the performance of the unlimited sampling scheme in real-world applications.

Conventional multi-bit ADCs require a very large number of quantization levels to represent the original continuous signal in high resolution settings. Sampling at high data rates with high resolution ADCs, however, would dramatically increase the overall power consumption and the manufacturing cost of such ADCs \cite{ameri2018one}. This problem is exacerbated in systems that require multiple ADCs such as large array receivers \cite{ho2019antithetic}. An immediate solution to such challenges is to use fewer bits for sampling. Therefore, in the recent years, the design of receivers with low-complexity \textit{one-bit ADC} has been emphasized to meet the requirements of both wide signal bandwidth and low cost/power. \emph{One-bit quantization} is an extreme quantization scenario, in which the ADCs are merely comparing the signals with given threshold levels, producing sign ($\pm1$) outputs. This enables the signal processing equipment to sample at a very high rate yet with considerably lower cost and energy consumption than the conventional ADCs \cite{instrumentsanalog,mezghani2018blind,eamaz2021modified,ameri2018one,sedighi2020one}.

\begin{comment}
\Ae{In high resolution settings, a very large number of quantization levels is required in order to represent the original continuous signal. However, this leads to some difficulties in applications where the signals of interest have large bandwidths, and may pass through several RF chains that require using a plethora of ADCs. Moreover, the large number of quantization bits can cause a considerable increase in the overall power consumption and the manufacturing cost of ADCs. Such drawbacks lend support to the idea of utilizing fewer bits for sampling. \emph{One-bit quantization} is an extreme quantization scenario, in which the ADCs are merely comparing the signals with given threshold levels, producing sign ($\pm1$) outputs. This enables the signal processing equipment to sample at a very high rate, with a considerably lower cost and energy consumption, compared to conventional ADCs \cite{instrumentsanalog,mezghani2018blind,eamaz2021modified,ameri2018one,sedighi2020one}}.
\end{comment}

In the classical problem of one-bit sampling, the signal is reconstructed by comparing the signal with a fixed, usually zero, threshold. This leads to difficulties in estimating signal parameters. In particular, when zero threshold is used, the power information of the input signal $\mathbf{x}$ is lost in one-bit data because the signs of $\mathbf{x}$ and $\eta\mathbf{x}$ are identical for $\eta>0$. This problem has been addressed in a few recent works \cite{eamaz2021modified,qian2017admm,gianelli2016one,eamaz2022phase,eamaz2022covariance,wang2017angular,xi2020gridless}, which show that time-varying sampling thresholds enable better estimation of the signal characteristics. In particular, time-varying thresholds were considered for the covariance recovery from one-bit measurements in \cite{eamaz2021modified}. This was extended  in \cite{eamaz2022covariance} for a significantly improved estimation of signal autocorrelation via the \emph{modified arcsine law}. In non-stationary scenarios, \cite{eamaz2022covariance} applied the modified arcsine law to utilize time-varying sampling thresholds. Applications of one-bit sampling to diverse problems such as sparse parameter estimation \cite{gianelli2016one}, localization \cite{sedighi2021localization}, and phase retrieval \cite{eamaz2022phase} have also appeared in the contemporary literature. %The neat idea to tackle the phase retrieval problem has been proposed in \cite{eamaz2022phase} which takes advantage of the abundance number of one-bit samples existing in the one-bit sampling with time-varying thresholds to make costly constraints such as positive semi-definite redundant. Moreover, time-varying sampling thresholds have been employed to recover signals of interest from the one-bit compressive sampling problem in \cite{gianelli2016one}.

%At the surface, one-bit sampling and unlimited sampling seem to address competing needs. 
Evidently, one-bit and unlimited sampling frameworks address complementary requirements. A one-bit ADC only compares an input signal with a given threshold. Therefore, essentially, one-bit sampling is indifferent to DR because, apart from the comparison bit, other information such as the distance between the signal value and the threshold is not stored. On the other hand, the self-reset ADC in unlimited sampling provides a natural approach to producing judicious time-varying thresholds for one-bit ADCs. In this paper, to harness advantages of both methods, we propose \emph{un}limited \emph{o}ne-bit (UNO) sampling to design sampling thresholds which are highly informative about the signal of interest. 

%\clearpage
\subsection{Prior Art}
\label{subsec:prior}
Unlimited sampling of continuous-time signals that are sparse in Fourier domain was discussed in \cite{bhandari2018unlimited}. Extensions to graph signals \cite{ji2019folded}, multi-channel arrays \cite{fernandez2021doa}, and sparse outliers (noise) \cite{bhandari2022unlimited} have also been proposed. Reconstruction algorithms have included wavelet-based \cite{rudresh2018wavelet}, generalized approximate message passing \cite{musa2018generalized}, and local average \cite{florescu2022unlimited} techniques. Very recently, non-idealities in hardware prototyping were considered in \cite{bhandari2021unlimited,beckmann2020hdr}; a computational sampling strategy in the form of \textit{unlimited sampling with hysteresis} \cite{florescu2021unlimited} was found to be more flexible for circuit design specifications. %Unlimited sampling based high dynamic range computed tomography was investigated in \cite{beckmann2020hdr}. 

To reconstruct the full-precision signal from one-bit sampled data, conventional approaches \cite{khobahi2018signal,zymnis2009compressed} include maximum likelihood estimation (MLE) and weighted least squares. However, these methods have high computational cost, especially for high-dimensional input signals. To this end, we propose using the randomized Kaczmarz algorithm (RKA) \cite{strohmer2009randomized,leventhal2010randomized}, which is an iterative algorithm for solving a system of linear inequalities that arise naturally in the one-bit quantization framework. While the deterministic version\cite{kaczmarz1937angenaherte} of the Kaczmarz method usually selects the linear equation sequentially, the RKA is random in its selection in each iteration leading to a faster convergence. The RKA is simple to implement and performs comparably with the state-of-the-art optimization methods. 

Among prior studies involving both one-bit and unlimited frameworks, state-of-the-art results in \cite{graf2019one} proposed \textit{one-bit $\Sigma\Delta$ quantization via unlimited sampling}, whose objective is to shrink the \textit{DR between the input signal and its one-bit samples}. This study developed a guaranteed reconstruction as long as the DR of the input signal is less than the DR of the one-bit data (i.e., 1). However, when the ratio of the input signal amplitude to the ADC threshold is large, then the imperfect noise shaping in sigma-delta conversion degrades this reconstruction. Contrary to this work, our proposed UNO technique focuses on a different problem, i.e., shrinking the \textit{DR between the input signal and the time-varying sampling thresholds}. The one-bit sampling is typically performed at significantly high rates. As a result, the resulting observation inequalities form an overdetermined system. %The UNO framework offers arbitrary time-varying sampling thresholds; thus, effectively reducing the gap between the input signal and the time-varying sampling thresholds. We show that the one-bit sampled signals are reconstructed efficiently using RKA.  Vijay: Keep all RKA discussion at one place. Already said this before
When the difference between the DR of the input signal and that of the thresholds increases, the reconstruction degrades significantly. We show that jointly exploiting both unlimited and one-bit sampling techniques provides a more efficient solution by a considerable reduction of the aforementioned gap. % between the input signal and the time-varying sampling thresholds. 
%In other words, unlimited sampling imposes a magnitude constraint on the signal and stores the information on the distance between the input signal values and the time-varying threshold. This knowledge becomes helpful in one-bit signal reconstruction.

%\textcolor{blue}{Instead of considering the often formulated maximum likelihood estimation (MLE) and relaxation problems, our approach to recover one-bit sampled data takes advantage of as a variant of the Kaczmarz algorithm.} \textcolor{red}{already said all of this two para before}

In practice, errors arising from quantization noise degrade the reconstruction quality in unlimited sampling framework.
%In practice, we observe that the original signal is contaminated with a noise which can come from round-off quantization error or the effect of environment. Due to the fact that the additive noise may change the nature of signal, the signal reconstruction task becomes more complicated which requires \emph{denoising} process to fully reconstruct the original signal.
In this context, \cite{bhandari2020unlimited} derived the reconstruction guarantees by including this error as bounded additive noise to the modulo samples.
%In \cite{bhandari2020unlimited}, they considered the effect of bounded noise on the reconstruction procedure of the input signal from modulo samples. Also, they proved that if the dynamic range of the additive noise (to modulo samples) satisfies a certain level, the unlimited sampling theorem guarantees the perfect reconstruction with negligible error. 
Contrary to this approach, we consider the more realistic case of additive noise to the input signal. %We present that, under certain condition on the additive noise, UNO-based reconstruction will be guaranteed. 
We show that %when noisy measurements are utilized, 
our RKA-based reconstruction is also effective for noisy one-bit sampled signals because it is independent from the statistical properties of the modulo samples. 
%\vspace{-8pt}
\subsection{Our Contributions}
\label{subsec:contrib}
%To guarantee the original signal reconstruction from UNO sampled data, a lower bound for the ADC thresholds is proposed which relies on the reconstruction error of the RKA. }
Our main contributions in this paper are:\\
\textbf{1) Combined unlimited and one-bit sampling framework.} In the proposed UNO framework, we leverage upon the benefits of both one-bit and unlimited sampling techniques. The result is a sampling approach that  yields unlimited DR and a low-cost,  low-power receiver while retaining a high sampling rate. We design time-varying sampling thresholds for one-bit quantization, whose DR is closer to that of the original signal. %By employing this threshold in one-bit sampling, 
This aids in accurately storing the information of distance between the signal values and thresholds to utilize in the signal reconstruction task. We show that compared to the one-bit reconstruction with random thresholds \cite{ameri2018one}, our proposed UNO sampling based on time-varying thresholds performs better, especially for high DR signals.
%Proposing a neat low-resolution sampling framework taking advantage of both unlimited sampling and one-bit quantization with time-varying sampling thresholds. 
\\
\textbf{2) RKA-based reconstruction.} The signal reconstruction from one-bit measurements requires solving an overdetermined linear feasibility problem that we recast as a one-bit polyhedron and efficiently solve it via the RKA. By generating an abundant number of one-bit samples, we show that the singular values of one-bit data matrix that creates the one-bit polyhedron are equal to the number of time-varying threshold sequences employed in one-bit sampling. Further, we numerically investigate the effects of ADC threshold and signal amplitude in the RKA-based UNO reconstruction. %By generating an abundant number of one-bit samples, which is a common scenario in the one-bit sampling with time-varying thresholds, the signal reconstruction problem turn to a highly overdetermined linear feasibility problem efficiently solved by the RKA. To verify the efficacy of the UNO-based signal reconstruction, we provide several numerical investigations at which we apply it algorithm to signals with different amplitudes and various ADC thresholds. Moreover, the effect of number of time-varying sampling threshold sequences on the proposed reconstruction algorithm is numerically investigated. 
\\
\textbf{3) Performance guarantees.} Our theoretical analyses show that a proper selection of the sufficient number of samples further enhances the reconstruction performance of the UNO. %Theoretically and numerically showing that the number of time-varying thresholds sequences is so effective in the UNO-based reconstruction performance.  
We prove that the convergence rate of the RKA when applied to the one-bit polyhedron depends on the size of the input signal and the total number of RKA iterations. In this context, we also obtain a lower bound on the number of required iterations for perfect reconstruction. %We show that the convergence rate of the proposed signal reconstruction solely depends on the signal dimension and the number of iterations employed in the RKA. Moreover, a lower bound is obtained for the total number of iterations required for the RKA in order to perfectly reconstruct the input signals. 
\\
    %\item Proving that to guarantee the perfect reconstruction of signal in the UNO, the ADC threshold must be higher than a lower bound depending on the RKA reconstruction error.
\textbf{4) Reconstruction in the presence of additive noise.} When the input signal is contaminated with additive noise, we apply the recently introduced new plug-and-play (PnP) priors \cite{venkatakrishnan2013plug} to the alternating direction method of multipliers (ADMM) as an additional reconstruction algorithm step. %to recover parameters of interest from the \emph{inverse problem}. 
In image denoising problems, the PnP-ADMM is used to replace the shrinkage step of the standard ADMM algorithm with any off-the-shelf algorithm %in the class of \emph{image denoisers} 
to ensure the noise variance is sufficiently suppressed. %An intersting observation regarding PnP-ADMM is that 
Although PnP-ADMM appears \textit{ad hoc}, it yields a better performance than state-of-the-art methods in several different inverse problems \cite{venkatakrishnan2013plug,chan2016plug}. For the noisy UNO, we deploy this algorithm to reconstruct the original signal from overdetermined and underdetermined noisy systems. Moreover, we show that the additive noise to the input signal contaminates the modulo samples with noise that is expressed in terms of the input noise.

%we propose Equipping the UNO with the PnP-ADMM algorithm to reconstruct . We show that the UNO-based reconstruction has good performance even when the measurements are noisy.

%\textcolor{red}{you should summarize in words what the main theorems of the paper show} \textcolor{blue}{to show the effect of the abundance sample-size regime,} \textcolor{red}{what does this mean?} In the presence of noise, .... \textcolor{red}{write about some prior works on noisy sampling, including classical results}. We show that, when noisy measurements are utilized, \textcolor{red}{our proposed UNO results in ....} %he performance of our approach is then investigated .  The reconstruction \textcolor{red}{check recovery vs reconstruction throughout the paper} based on Kaczmarz algorithm is also effective for %recover the 
%noisy one-bit sampled signals % is studied and illustrated.
%\textcolor{red}{because ....}

%\textcolor{red}{so many contributions have been omitted form this subsection: error bound, numerical results, etc.}
%\Ae{We can summarize our contributions in this paper as follows: }

%\textcolor{red}{Preliminary results of this work appeared in our conference publication \textcolor{blue}{cite icassp paper} %\cite{eamaz2023icassp}, 
%where only ..... was examined. In this paper, we further investigate and do a, b, and c....} \Fr{theoretical study on number of threshold sequences, noisy case, theoretical study on the value of $\lambda$ for noisy case.}.
%\vspace{-8pt}
\subsection{Organization and Notations} %s of the Paper}
In the next section, %Section~\rom{2} is dedicated to a 
we provide an introduction to one-bit quantization with time-varying sampling thresholds. Particularly, the one-bit sampled signal reconstruction problem is formulated as an overdetermined system of linear inequalities. In Section~\ref{sec:unlimited}, we recall the concept of unlimited sampling as proposed in \cite{bhandari2017unlimited,bhandari2020unlimited}. We introduce the RKA in the context of signal reconstruction in Section~\ref{sec:onebit_rec}. %to solve linear inequality constraints (defining a polyhedron feasible region), and to recover the desired signal.
%The recovery error and a roadmap to select the number of iterations of the RKA is also discussed in Section~\rom{4}. 
This is a prelude to Section~\ref{sec:uno}, which proposes UNO sampling to design judicious thresholds and guarantee the one-bit signal reconstruction in the high-DR. In Section~\ref{numerical_unlim}, we provide several numerical experiments to illustrate UNO-based sampling and analyze the reconstruction error. We consider the noisy measurement scenario in Section~\ref{sec:noise} %is studied owing to its importance in practical settings. We 
and conclude in Section~\ref{sec:summ}.

%\Fr{It's better to keep the name of the notation in the journal paper because we do it in each of our journal paper.} \textcolor{red}{yes. we already included `Notations' in the section heading. It is the same thing.}
%\underline{\emph{Notation:}}
Throughout this paper, we use boldface lowercase, boldface uppercase, and calligraphic letters for vectors, matrices, and sets, respectively. The notations $\mathbb{C}$, $\mathbb{R}$, and $\mathbb{Z}$ represent the set of complex, real, and integer numbers, respectively. We represent a vector $\mathbf{x}$ in terms of its elements $\{x_{i}\}$ or $\left(\mathbf{x}\right)_{i}$ as $\mathbf{x}=[x_{i}]$. %\textcolor{red}{notational conflict. Here, you are using $x_{i}$, $\{x_{i}\}$, and $\left(\mathbf{x}\right)_{i}$ for the elements of a vector. But later, you use a similar subscript notation for the $i$-th iteration of vector, i.e.,  $\mathbf{x}_{i}$ in Section IV.B. Why do we need so many notations for the i-th element. Use just one notation. Also, $\{x_{i}\}$ is the sequence or collection of elements, not the i-th element itself. But the text here indicates differently. For iterations, you could have simply used $\mathbf{x}^{(i)}$ in Section IV.B and other places. That is the standard iteration notation (ie put the iteration number in parentheses in the superscript)} \Fr{But, again we will have a notation conflict with $\mathbf{r}^{(\ell)}$ and $\boldsymbol{\uptau}^{(\ell)}$. For samples, we use $x_{i}$ or $\left(\mathbf{x}\right)_{i}$ as the $i$-th element of the vector $\mathbf{x}$. For an iterative process, $\mathbf{x}_{i}$ denotes the vector $\mathbf{x}$ in the $i$-th iteration. I think it is clear during the paper.} \textcolor{red}{need to resolve the notational conflict. reviewers will easily pick on this} \Fr{I still did not get the notation conflict!} 
We use $(\cdot)^{\top}$ and $(\cdot)^{\mathrm{H}}$ to denote the vector/matrix transpose and the Hermitian transpose, respectively. The identity matrix of size $N$ is $\mathbf{I}_{N}\in \mathbb{R}^{N\times N}$. The Frobenius norm of a matrix $\mathbf{B}\in \mathbb{C}^{M\times N}$ is defined as $\|\mathbf{B}\|_{\mathrm{F}}=\sqrt{\sum^{M}_{r=1}\sum^{N}_{s=1}\left|b_{rs}\right|^{2}}$, where $b_{rs}$ is the $(r,s)$-th entry of $\mathbf{B}$. The function $\text{diag}(\cdot)$ outputs a diagonal matrix with the input vector along its main diagonal. The $\ell_{p}$-norm of a vector $\mathbf{b}$ is $\|\bb\|_{p}=\left(\sum_{i}b^{p}_{i}\right)^{1/p}$. The infinity or max-norm %($p\rightarrow \infty$) 
of a function $x$ is  $\|x\|_{\infty}=\operatorname{inf}\left\{c_{0}\geq 0: |x(t)|\leq c_{0}\right\}$, where $\textrm{inf}(\cdot)$ denotes the infimum of its argument; for vectors, we have $\|\mathbf{x}\|_{\infty}=\max_{k}|x_{k}|$. For a vector $\mathbf{x}$, $\Delta\mathbf{x}=x_{k+1}-x_{k}$ denotes the finite difference and recursively applying the same yields $N$-th order difference, $\Delta^{N}\mathbf{x}$. We denote the $\Omega$-bandlimited Paley-Wiener subspace of the square-integrable function space $L^{2}$ by $\textrm{PW}_{\Omega}$ such that $\textrm{PW}_{\Omega}=\{f: f, \widehat{f} \in L^2,\;\operatorname{supp}(\widehat{f}) \subset[-\Omega, \Omega]\}$, where $\widehat{f}$ is the Fourier transform of $f$. The Hadamard (element-wise) product of two matrices $\mathbf{B}_{1}$ and $\mathbf{B}_{2}$ is $\mathbf{B}_{1}\odot \mathbf{B}_{2}$. The column-wise vectorized form of a matrix $\mathbf{B}$ is  $\operatorname{vec}(\mathbf{B})$. Given a scalar $x$, we define the operator $(x)^{+}$ as $\max\left\{x,0\right\}$. For an event $\mathcal{E}$, $\mathbb{1}_{(\mathcal{E})}$ is the indicator function for that event meaning that $\mathbb{1}_{(\mathcal{E})}$ is $1$ if $\mathcal{E}$ occurs; otherwise, it is zero. The function $\operatorname{sgn}(\cdot)$ yields the sign of its argument. In the context of numerical computations, $\lfloor\rfloor$ and $\lceil\rceil$ denote the floor and ceiling functions, respectively. The function $\log(\cdot)$ denotes the natural logarithm, unless its base is otherwise stated. The notation $x \sim \mathcal{U}(a,b)$ means a random variable drawn from the uniform distribution over the interval $[a,b]$ and $x \sim \mathcal{N}(\mu,\sigma^2)$ represents the normal distribution with mean $\mu$ and variance $\sigma^2$. The operator $\operatorname{mod}(a,b)$ between two values $a$ and $b$, returns the remainder of the division operation $a/b$.
\begin{comment}
The cumulative distribution function (CDF) of the zero-mean truncated Gaussian process $z\sim\mathcal{N}(0,\zeta)$ is given by
\begin{equation}
\label{eq:1bbb}
\Phi(z) \triangleq \frac{1}{\sqrt{2\pi}}\int^{z}_{-\infty}e^{-\frac{t^{2}}{2\zeta^{2}}} \,dt.
\end{equation}
The CDF of the truncated Gaussian process $\Tilde{z}\sim\mathcal{TN}(\mu,\sigma_{\Tilde{z}},\alpha,\beta)$ is given by
\begin{equation}
\label{eq:1bbbb}
\begin{aligned}
F(\Tilde{z}) &\triangleq \frac{\Phi\left(\frac{\Tilde{z}-\mu}{\sigma_{\Tilde{z}}}\right)-\Phi\left(\frac{\alpha-\mu}{\sigma_{\Tilde{z}}}\right)}{\chi},\\
\chi &= \Phi\left(\frac{\beta-\mu}{\sigma_{\Tilde{z}}}\right)-\Phi\left(\frac{\alpha-\mu}{\sigma_{\Tilde{z}}}\right),
\end{aligned}
\end{equation}
where $\Tilde{z}\in \left[\alpha, \beta\right]$, and $\Phi(.)$ is the CDF of a zero-mean Gaussian process. Finally, the error function (erf) is defined as $
\operatorname{erf} x=\frac{2}{\sqrt{\pi}} \int_{0}^{x} e^{-z^{2}}\,d z$.
\end{comment}

%\vspace{-8pt}
\section{One-Bit Sampling: Overdetermined Linear System Formulation}
\label{sec:onebit}
%\textcolor{blue}{In this section, at first we will present a review of one-bit sampling via time-varying thresholds. Then, we move to formulate this sampling structure based on an overdetermined linear system of inequalities.} \textcolor{red}{already said all of this in the outline. No need to say this again. It is not considered good writing. Use this leading para to describe some other connections with the literature. As an example, see how I write the leading paragraphs before the subsections in each section of this paper: \url{https://www.researchgate.net/publication/357578952_Inverse_Extended_Kalman_Filter_---_Part_I_Fundamentals}. Such discussions are important and show depth of the work.}\Ae{I cannot clearly get your point here. I added complete literature review and our contributions in the introduction. What should I do exactly here except a brief review of this section?}
%\textcolor{red}{write this para like this. fill in or substitute the details accordingly.} 
Several approaches have been proposed in the literature to reconstruct the signal of interest from one-bit samples with the most of them formulating this task as an optimization problem. For example, the covariance matrix formulation of \cite{ameri2018one} employs the cyclic optimization method to recover the input autocorrelation elements. A convex program based on the Gauss-Legendre integration to recover the input covariance matrix from one-bit sampled data was suggested in \cite{eamaz2022covariance}. Other recent works exploit sparsity of the signal and apply techniques such as $\ell_1$-norm minimization \cite{zahabi2020one,knudson2016one}, $\ell_{1}$-regularized MLE formulation \cite{zymnis2009compressed,khobahi2019model}, log-relaxation \cite{zhu2020target}, and Lasserre's semidefinite program relaxation \cite{sedighi2021localization} to lay the ground for signal reconstruction. In the following, we explain our one-bit polyhedron formulation, wherein a strong efficient and easily implementable solver of linear feasibility problems is applied to the aforementioned application-specific methods.
%\vspace{-8pt}
\subsection{One-Bit Quantization Using Time-Varying Thresholds}
\label{sec:survey}
Consider a bandlimited continuous-time signal $x\in \textrm{PW}_{\Omega}$
that we represent via Shannon's sampling theorem as \cite{hogan2012duration}
\begin{equation}
\label{eq:1}
0<\mathrm{T} \leqslant \frac{\pi}{\Omega}, \quad x(t)=\sum_{k=-\infty}^{k=+\infty} x(k \mathrm{T}) \operatorname{sinc}\left(\frac{t}{\mathrm{T}}-k\right),
\end{equation}
where $1/\mathrm{T}$ is the sampling rate, $\Omega$ is the signal bandwidth, and $\operatorname{sinc}(t)=\frac{\sin (\pi t)}{\pi t}$ is an \emph{ideal} low-pass filter. Denote the uniform samples of $x(t)$ with the sampling rate $1/\mathrm{T}$ by $x_{k}=x(k\mathrm{T})$.

In practice, the discrete-time samples occupy pre-determined quantized values. We denote the quantization operation on $x[k]$ by the function $Q(\cdot)$. This yields the quantized signal as $r_{k} = Q(x_{k})$.
\begin{comment}
\begin{equation}
\label{eq:2}
r_{k} = Q(x_{k}),
\end{equation}
\end{comment}
In one-bit quantization, compared to zero or constant thresholds, time-varying sampling thresholds yield a better reconstruction performance \cite{ameri2018one,eamaz2022covariance}. These thresholds may be chosen from any distribution. In this work, to be consistent with state-of-the-art \cite{ameri2018one,khobahi2018signal,eamaz2021modified}, we consider a Gaussian non-zero time-varying threshold vector $\boldsymbol{\uptau}=\left[\tau_{k}\right]$ that follows the distribution $\boldsymbol{\uptau} \sim \mathcal{N}\left(\mathbf{d}=\mathbf{1}d,\bSigma\right)$. For one-bit quantization with such time-varying sampling thresholds, $r_{k} = \operatorname{sgn}\left(x_{k}-\tau_{k}\right)$.
\begin{comment}
\begin{equation}
\label{eq:3}
r_{k} = \operatorname{sgn}\left(x_{k}-\tau_{k}\right).
\end{equation}
\end{comment}
%\vspace{-8pt}
\subsection{One-Bit Polyhedron}
\label{subsec:overdetermined}
The information gathered through the one-bit sampling with time-varying thresholds may be formulated in terms of an overdetermined linear system of inequalities. We have $r_{k}=+1$ when $x_{k}>\tau_{k}$ and $r_{k}=-1$ when $x_{k}<\tau_{k}$. Collecting all the elements in the vectors as $\mathbf{x}=[x_{k}] \in \mathbb{R}^{n}$ and $\mathbf{r}=[r_{k}] \in \mathbb{R}^{n}$, therefore, one can formulate the geometric location of the signal as 
\begin{equation}
\label{eq:4}
r_{k}\left(x_{k}-\tau_{k}\right) \geq 0.
\end{equation}
Then, the vectorized representation of (\ref{eq:4}) is $\mathbf{r} \odot \left(\mathbf{x}-\boldsymbol{\uptau}\right) \geq \mathbf{0}$
\begin{comment}
\begin{equation}
\label{eq:5}
\mathbf{r} \odot \left(\mathbf{x}-\boldsymbol{\uptau}\right) \geq \mathbf{0},
\end{equation}
\end{comment}
or equivalently
\begin{equation}
\label{eq:6}
\begin{aligned}
\bOmega \mathbf{x} &\succeq \mathbf{r} \odot \boldsymbol{\uptau},
\end{aligned}
\end{equation}
where $\bOmega \triangleq \operatorname{diag}\left(\mathbf{r}\right)$. Suppose $\mathbf{x},\boldsymbol{\uptau} \in \mathbb{R}^{n}$, and that $\boldsymbol{\uptau}^{(\ell)}$ denotes the time-varying sampling threshold in $\ell$-th signal sequence, where  $\ell\in\mathcal{L}=\{1,\cdots,m\}$. 

For the $\ell$-th signal sequence, (\ref{eq:6}) becomes
\begin{equation}
\label{eq:7}
\begin{aligned}
\bOmega^{(\ell)} \mathbf{x} &\succeq \mathbf{r}^{(\ell)} \odot \boldsymbol{\uptau}^{(\ell)}, \quad \ell \in \mathcal{L},
\end{aligned}
\end{equation}
where $\bOmega^{(\ell)}=\operatorname{diag}\left(\mathbf{r}^{(\ell)}\right)$. Denote the concatenation of all $m$ sign matrices as 
\begin{equation}
\label{eq:9}
\Tilde{\bOmega}=\left[\begin{array}{c|c|c}
\bOmega^{(1)} &\cdots &\bOmega^{(m)}
\end{array}\right]^{\top}, \quad %\Tilde{\bOmega}
\in \mathbb{R}^{m n\times n}.
\end{equation}
Rewrite the $m$ linear system of inequalities in  (\ref{eq:7}) as
\begin{equation}
\label{eq:8}
\Tilde{\bOmega} \mathbf{x} \succeq \operatorname{vec}\left(\mathbf{R}\right)\odot \operatorname{vec}\left(\bGamma\right),
\end{equation}
where $\mathbf{R}$ and $\bGamma$ are matrices, whose columns are the sequences $\left\{\mathbf{r}^{(\ell)}\right\}_{\ell=1}^{m}$ and $\left\{\boldsymbol{\uptau}^{(\ell)}\right\}_{\ell=1}^{m}$, respectively. % which $\left\{\mathbf{r}^{(\ell)}\right\}_{\ell=1}^{m}$ and $\left\{\mathbf{\uptau}^{(\ell)}\right\}_{\ell=1}^{m}$ are sequences that are assigned to the columns of $\mathbf{R}$ and $\bGamma$, respectively,} and $\Tilde{\bOmega}$ is

%Hereafter, we consider (\ref{eq:8}) as an overdetermined linear system of inequalities associated with the one-bit sampling scheme presented in (\ref{eq:3}). The inequality (\ref{eq:8}) can be recast by a polyhedron as
The linear system of inequalities in (\ref{eq:8}) associated with the one-bit sampling scheme is overdetermined. We recast (\ref{eq:8}) into a \textit{one-bit polyhedron} as
\begin{equation}
\label{eq:8n}
\begin{aligned}
\mathcal{P} = \left\{\mathbf{x} \mid \Tilde{\bOmega} \mathbf{x} \succeq \operatorname{vec}\left(\mathbf{R}\right)\odot \operatorname{vec}\left(\bGamma\right)\right\}.
\end{aligned}
\end{equation}
%which we refer to as the \emph{one-bit polyhedron}. 
Instead of complex high-dimensional optimization with techniques such as MLE, our objective is to employ the polyhedron (\ref{eq:8n}) that encapsulates the desired signal $\mathbf{x}$ and leads to solving linear inequalities with linear convergence in expectation.
%\vspace{-8pt}
%\textcolor{red}{COMPLETED edits until here}
\section{Unlimited Sampling}
\label{sec:unlimited}
%Herein, we present a review of the unlimited sampling proposed in \cite{bhandari2017unlimited,bhandari2020unlimited}. 
In a variety of applications, clipping or saturation poses a serious problem to signal reconstruction. For instance, in scientific imaging systems such as ultrasound \cite{olofsson2005deconvolution}, radar \cite{cassidy2009ground}, and seismic imaging \cite{zhang2016restoration}, strong reflections or pulse echoes blind the sensor. In audio processing, clipped sound results in high-frequency artifacts \cite{adler2011audio}. In this context, unlimited sampling suggests that, instead of point-wise samples of the bandlimited function $x(t)$, the signal is digitized using a self-reset ADC with an appropriately selected threshold  $\lambda>0$ such that any signal value outside the range $\left[-\lambda,\lambda\right]$ is \emph{folded} to the same range \cite{bhandari2017unlimited,bhandari2020unlimited}. The folding corresponds to introducing a non-linearity in the sensing process \cite{bhandari2017unlimited,bhandari2020unlimited}. We denote the folding by the modulo operator $\mathcal{M}_{\lambda}$ that represents the following mapping:
\begin{equation}
\label{eq:18}
\mathcal{M}_{\lambda}(x_{k}): \Tilde{x}_{k} = x_{k}-2\lambda \left\lfloor\frac{x_{k}}{2\lambda}+\frac{1}{2}\right\rfloor,
\end{equation}
where $\Tilde{x}_{k}$ are the modulo samples of $x(t)$. 

The \emph{unlimited sampling theorem} \cite{bhandari2020unlimited} (reproduced below) states that, if the estimate of the norm of the bandlimited signal is known, then its perfect reconstruction (up to additive multiples of $2\lambda$) from its modulo samples is possible with at least sampling period $T \leq(2 \pi e)^{-1}$, where $e$ is the Euler's number and the signal bandwidth has been normalized to $\pi$.
\begin{theorem}[Unlimited sampling theorem \cite{bhandari2020unlimited}]
\label{theorem_1}
Assume $x(t)$ to be a finite energy, bandlimited signal with maximum frequency $\Omega_{\textrm{max}}$ and let $\Tilde{x}_{k}$, $k \in \mathbb{Z}$ in (\ref{eq:18}) be the modulo samples of $x(t)$ with sampling rate $1/T$. Then a sufficient condition for the reconstruction of $x(t)$ from $\left\{\Tilde{x}_{k}\right\}$ is that $T \leq \frac{1}{2\Omega_{\textrm{max}} e}$ (up to additive multiples of $2\lambda$).
\end{theorem}
Theorem~\ref{theorem_1} implies that the sampling rate depends on only the bandwidth and is independent of the ratio of ADC threshold $\lambda$ to the signal amplitude. In other words, the DR of the input signal is \emph{unlimited}. Recently, stable unlimited sampling reconstruction in the presence of noise has also been obtained \cite{bhandari2020unlimited}. %We recall this result later in Section~\ref{sec:noise}.

%Based on the mathematical formalism of the modulo operator $\mathcal{M}_{\lambda}$ in (\ref{eq:18}), the following remark is the stepping stone towards the reconstruction of the bandlimited function $x(t)$ from its modulo samples $\left\{\Tilde{x}_{k}\right\}$ \cite{bhandari2017unlimited,bhandari2020unlimited}. 
%\underline{\emph{Remark}:}
%\label{prop_1}
%Let $x$ be a bandlimited function and $\mathcal{M}_{\lambda}$ denote the modulo operator defined in (\ref{eq:18}), where $\lambda$ is a fixed, positive constant. Then, $x(t)$ admits a decomposition \cite{bhandari2017unlimited,bhandari2020unlimited},
%\begin{equation}
%\label{eq:19}
%x(t) = \Tilde{x}(t) + \epsilon_{x}(t),
%\end{equation}
%where $\Tilde{x}(t)=\mathcal{M}_{\lambda}\left(x(t)\right)$ and $\epsilon_{x}$ is
%\begin{equation}
%\label{eq:20}
%\epsilon_{x}(t)=2\lambda \sum_{u \in \mathbb{Z}}e_{u}\mathbb{1}_{\mathcal{D}_{u}}(t), \quad e_{u} \in \mathbb{Z},
%\end{equation}
%where $\bigcup_{u \in \mathbb{Z}} \mathcal{D}_{u}=\mathbb{R}$ is a partition of the real line into intervals $\mathcal{D}_{u}$.

The reconstruction of the bandlimited function $x(t)$ from its modulo samples $\left\{\Tilde{x}_{k}\right\}$ is achieved as follows. Assume that $x(t)$ admits a decomposition \cite{bhandari2017unlimited,bhandari2020unlimited}, 
\begin{equation}
\label{eq:19}
x(t) = \Tilde{x}(t) + \epsilon_{x}(t),
\end{equation}
where $\Tilde{x}(t)=\mathcal{M}_{\lambda}\left(x(t)\right)$ and the error $\epsilon_{x}$ between the input signal and its modulo samples is
\begin{equation}
\label{eq:20}
\epsilon_{x}(t)=2\lambda \sum_{u \in \mathbb{Z}}e_{u}\mathbb{1}_{\mathcal{D}_{u}}(t), \quad e_{u} \in \mathbb{Z},
\end{equation}
where $\bigcup_{u \in \mathbb{Z}} \mathcal{D}_{u}=\mathbb{R}$ is a partition of the real line into intervals $\mathcal{D}_{u}$. As indicated by (\ref{eq:19}), if $\epsilon_{x}$ is known, then $x$ can be reconstructed from $\Tilde{x}$. It follows from \eqref{eq:20} that $\epsilon_{x}$ takes only those values that are integer multiples of $2\lambda$ thereby leading to a robust reconstruction algorithm \cite{bhandari2020unlimited}. To obtain $\epsilon_{x}$ (up to an unknown additive constant) and subsequently the desired signal $x(t)$, the reconstruction procedure in \cite{bhandari2017unlimited,bhandari2020unlimited} requires the higher-order differences of $\Tilde{\mathbf{x}}=[\Tilde{x}_{k}]$ to obtain $\Delta^{N}\bepsilon_{x}=\mathcal{M}_{\lambda}\left(\Delta^{N}\Tilde{\mathbf{x}}\right)-\Delta^{N}\Tilde{\mathbf{x}}$, where $\bepsilon_{x}=[\epsilon_{x}]$. Define the inverse-difference operator as a sum of real sequence $\{s_b\}$, 
i.e.,
\begin{equation}
\label{eq:23}
\nabla: \{s_{k}\}_{k \in \mathbb{Z}^{+}} \rightarrow \sum_{b=1}^{k}s_{b}.
\end{equation}
Then, applying $\nabla\left(\Delta^{N}\bepsilon_{x}\right)$ and rounding the result to the nearest multiple of $2\lambda \mathbb{Z}$ yields $\epsilon_{x}$. For a guaranteed and stable reconstruction performance, a suitable choice for difference order $N$ is \cite{bhandari2020unlimited},
\begin{equation}
\label{eq:21}
N \geq \left\lceil \frac{\log \lambda -\log \beta_{x}}{\log\left(T\Omega e\right)} \right\rceil,
\end{equation}
where $\beta_{x}$ is chosen such that
$\beta_{x} \in 2\lambda \mathbb{Z}$ and $\|x\|_{\infty} \leq \beta_{x}$.
Algorithm~\ref{algorithm_1} summarizes the unlimited sampling reconstruction procedure.

%-------------------------------------------------
\begin{algorithm}[H] %Vijay: Always use H option for algorithms, not t. Algorithms are part of the text and come right after you refer to them. They are not figures or tables.
	\caption{Input signal reconstruction from modulo folded samples.}
    \label{algorithm_1}
    \begin{algorithmic}[1]
    \Statex \textbf{Input:} $\Tilde{x}_{k}=\mathcal{M}_{\lambda}\left(x_{k}\right)$, ADC threshold $\lambda$, and $2\lambda \mathbb{Z}\ni\beta_{x}\geq \|x\|_{\infty}$. 
    \Statex \textbf{Output:}  The approximation of the input signal $\bar{\mathbf{x}}$. %\textcolor{red}{reconstructed signal uses hat, which you also use for Fourier transform. In the text, you use barx for the reconstructed signal} \Fr{Done.}

    \State %Compute 
    $N \gets \left\lceil \frac{\log \lambda -\log \beta_{x}}{\log\left(T\Omega e\right)} \right\rceil$ using (\ref{eq:21}).

    \State %Compute
    $\Delta^{N}\bepsilon_{x} \gets \mathcal{M}_{\lambda}\left(\Delta^{N}\Tilde{\mathbf{x}}\right)-\Delta^{N}\Tilde{\mathbf{x}}$.

    \State %Set
    $\mathbf{s}_{0}\gets\Delta^{N}\bepsilon_{x}$.
    
    \For{$p=0:N-2$} 
    \State $\mathbf{s}_{p+1}\gets\nabla\mathbf{s}_{p}$ $\triangleright$ $\nabla$ is the inverse-difference operator defined in (\ref{eq:23}).
    
    \State $\mathbf{s}_{p+1}\gets2\lambda\left\lceil\frac{\left\lfloor\mathbf{s}_{p+1}/\lambda\right\rfloor}{2}\right\rceil$ $\triangleright$ rounding to $2\lambda\mathbb{Z}$.
    
    \State %Compute
    $\kappa_{p}\gets\left\lfloor\frac{\left(\nabla^{2}\Delta^{p}\bepsilon_{x}\right)_{1}-\left(\nabla^{2}\Delta^{p}\bepsilon_{x}\right)_{J+1}}{12\beta_{x}}+\frac{1}{2}\right\rfloor$ $\triangleright$ $J=\frac{6\beta_{x}}{\lambda}$.
    
    \State $\mathbf{s}_{p+1}\gets\mathbf{s}_{p+1}+2\lambda\kappa_{p}$.
    \EndFor
    
    \State \Return $\bar{\mathbf{x}} \gets \nabla\mathbf{s}_{N-1}+\Tilde{\mathbf{x}}+2a\lambda, \quad a \in \mathbb{Z}$. 
    \end{algorithmic}
\end{algorithm}
%-------------------------------------------------

%\vspace{-8pt}
\section{One-Bit Signal Reconstruction}
\label{sec:onebit_rec}
%The goal of one-bit signal reconstruction is to estimate the unknown signal $\mathbf{x}$ from one-bit measurements $\left\{\mathbf{r}^{(\ell)}\right\}_{\ell=1}^{m}$ using the set of inequalities in (\ref{eq:8n}). 
%For this purpose, we use the \emph{randomized Kaczmarz algorithm} (RKA) which will be introduced in Subsection~\ref{RKA}. Then, we move to formulate the convergence rate of the proposed algorithm by obtaining an error reconstruction boundary. In Subsection~\ref{NUM_RKA}, the reconstruction performance of RKA will be numerically evaluated. Finally, the challenges faced by RKA in signal reconstruction will be investigated in Section~\ref{Guaranteed}.
To reconstruct %the signal of interest  
$\mathbf{x}$ from the sign data $\left\{\mathbf{r}^{(\ell)}\right\}^{m}_{\ell=1}$, we %consider the overdetermined linear system of inequalities in (\ref{eq:8n}). We 
solve the polyhedron search problem through RKA because of its optimal projection and linear convergence in expectation \cite{eamaz2022phase,briskman2015block,leventhal2010randomized}.
%\vspace{-8pt}
\subsection{Basic Theory of RKA}
\label{RKA}
The RKA is a \emph{subconjugate gradient method} to solve overdetermined linear systems, i.e, $\mathbf{C}\mathbf{x}\leq\mathbf{b}$ where $\mathbf{C}$ is a $m^{\prime}\times n^{\prime}$ matrix with $m^{\prime}>n^{\prime}$ \cite{leventhal2010randomized,strohmer2009randomized}. The conjugate-gradient methods turn this inequality to an equality of the following form:
\begin{equation}
\label{eq:10}
\left(\mathbf{C}\mathbf{x}-\mathbf{b}\right)^{+}=0,
\end{equation}
and then solve it as any other system of equations. Given a sample index set $\mathcal{J}$, without loss of generality, rewrite (\ref{eq:10}) as the polyhedron
\begin{equation}
\label{eq:11}
\begin{aligned}
\begin{cases}\mathbf{c}_{j} \mathbf{x} \leq b_{j} & \left(j \in \mathcal{I}_{\leq}\right), \\ \mathbf{c}_{j} \mathbf{x}=b_{j} & \left(j \in \mathcal{I}_{=}\right),\end{cases}
\end{aligned}
\end{equation}
where $\{\mathbf{c}_{j}\}$ are the rows of $\mathbf{C}$ and the disjoint index sets $\mathcal{I}_{\leq}$ and $\mathcal{I}_{=}$ partition $\mathcal{J}$. The projection coefficient $\beta_{i}$ of the RKA is \cite{leventhal2010randomized,briskman2015block,dai2013randomized}:
\begin{equation}
\label{eq:12}
\beta_{i}= \begin{cases}\left(\mathbf{c}_{j} \mathbf{x}_{i}-b_{j}\right)^{+} & \left(j \in \mathcal{I}_{\leq}\right), \\ \mathbf{c}_{j} \mathbf{x}_{i}-b_{j} & \left(j \in \mathcal{I}_{=}\right).\end{cases}
\end{equation}
The unknown column vector $\mathbf{x}$ is iteratively updated as
\begin{equation}
\label{eq:13}
\mathbf{x}_{i+1}=\mathbf{x}_{i}-\frac{\beta_{i}}{\left\|\mathbf{c}_{j}\right\|^{2}_{2}} \mathbf{c}^{\mathrm{H}}_{j},
\end{equation}
where, at each iteration $i$, the index $j$ is drawn from the set $\mathcal{J}$ independently at random following the distribution
\begin{equation}
\label{eq:14}
\operatorname{Pr}\{j=k\}=\frac{\left\|\mathbf{c}_{k}\right\|^{2}_{2}}{\|\mathbf{C}\|_{\mathrm{F}}^{2}}.
\end{equation}

Note that, (\ref{eq:8n}) has only the inequality partition $I_{\leq}$. Herein, $m^{\prime}=m\times n$ and $n^{\prime}=n$. The row vector $\mathbf{c}_{j}$ and the scalar $b_{j}$ in the RKA (\ref{eq:11})-(\ref{eq:14}) are $j$-th row of $-\Tilde{\bOmega}$ and $j$-th element of $-\left(\operatorname{vec}\left(\mathbf{R}\right)\odot\operatorname{vec}\left(\bGamma\right)\right)$, respectively. It may be readily verified that the distribution of choosing a specific sample index $j$ for the inequalities in (\ref{eq:8n}) is uniform, i.e., $\operatorname{Pr}\{j=k\}=\frac{1}{mn}$.

In one-bit reconstruction, $\mbc_{j}=-\upomega_{j}$, wherein $\upomega_{j}$ is the $j$-th row of $\Tilde{\bOmega}$; a $j^{\prime}$-th \emph{coordinate vector} with $\pm1$ as its $j^{\prime}$-th element and %in the $j^{\prime}$-th place where $j^{\prime}$ is 
\begin{equation}
\label{Negindex} 
\begin{aligned}
j^{\prime}=\begin{cases} \operatorname{mod}(j,n), & j\neq k n, \\ n, & j=k n,\end{cases}
\end{aligned}
\end{equation}
with $1\leq k \leq m$. This property makes the update process \eqref{eq:13} similar to that of the \emph{randomized Gauss-Seidal} method using the coordinate vector in each iteration \cite{leventhal2010randomized,ma2015convergence}. This approach is commonly used for solving high-dimensional linear feasibility problems %when the dimensions are extremely large, since at each iteration only a single dimension is updated. 
by updating only one dimension at any iteration. The structure of matrix $\Tilde{\bOmega}$ leads to a similar efficient RKA implementation %appears to be so efficient for high-dimensional input signals 
by updating only the generic element $j^{\prime}$  at each iteration, i.e., 
%\begin{equation}
%\label{Negindex2}
$\left(\mathbf{x}_{i+1}\right)_{j^{\prime}}=\left(\mathbf{x}_{i}\right)_{j^{\prime}}-\beta_{i} r_{j^{\prime}}$, 
%\end{equation}
where $r_{j^{\prime}}$ is the one-bit data at index $j^{\prime}$.

%----------------------------------------------------------------------
\begin{figure*}
	\centering
	\includegraphics[width=1.0\textwidth]{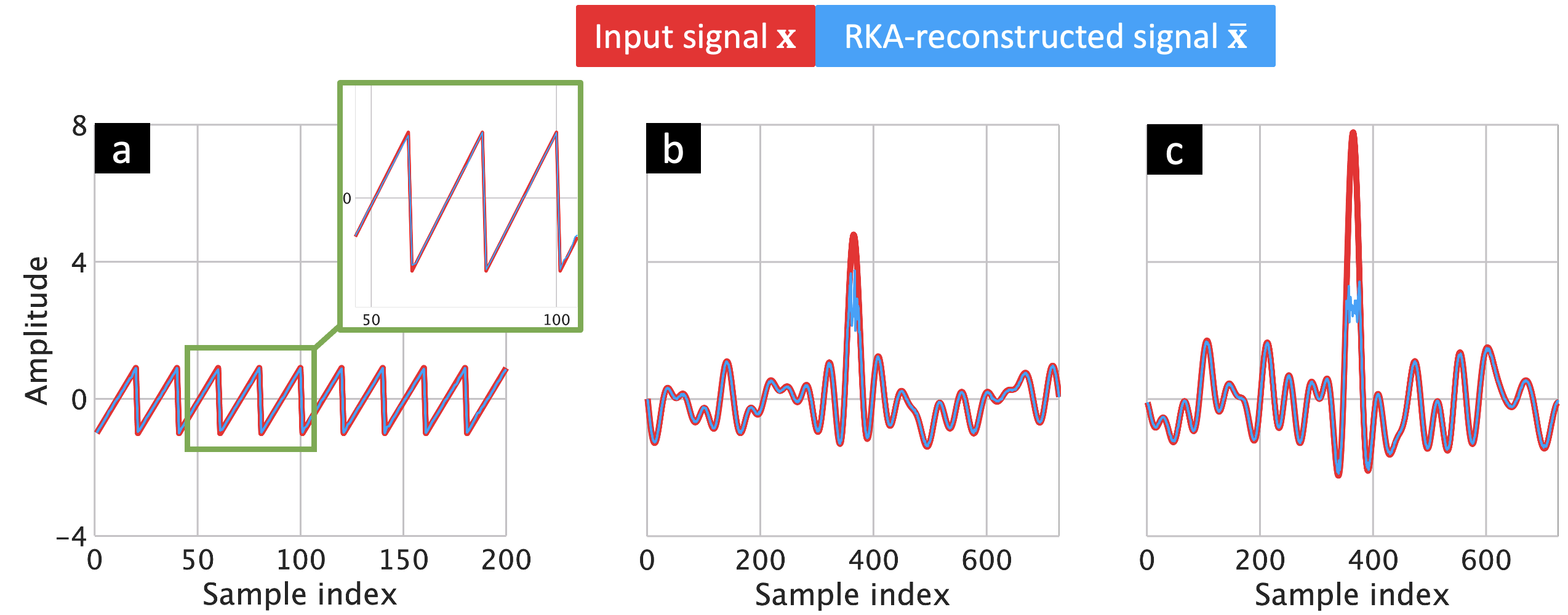}
	\caption{ (a) The input sawtooth wave signal $\mathbf{x}$ is reconstructed from one-bit measurements using the RKA to yield $\bar{\mathbf{x}}$. Here, $\text{DR}_{\mbx}=1$ and $\text{DR}_{\boldsymbol{\uptau}}=1$. The inset shows the same plot on a larger scale. (b) As in (a) but for the bandlimited input signal from \cite{bhandari2020unlimited} with $\text{DR}_{\mbx}=5$. 
	(c) As in (b) but for $\text{DR}_{\mbx}=8$.
    %\vspace{-10pt}
    }
	\label{figure_1}
\end{figure*}
%----------------------------------------------------------------------
%\vspace{-8pt}
\subsection{Error Reconstruction Bound}
\label{ERROR_BOUND}
At the $i$-th iteration, the error between the RKA estimate $\mathbf{x}_{i}$ and the optimal solution $\mathbf{x}^{\star}$ has been shown to follow the convergence bound \cite{strohmer2009randomized,leventhal2010randomized,briskman2015block,polyak1964gradient}
\begin{equation}
\label{eq:15}
\mathbb{E}\left\{\left\|\mathbf{x}_{i}-\mathbf{x}^{\star}\right\|_{2}^{2}\right\} \leq q^{i}\left\|\mathbf{x}_{0}-\mathbf{x}^{\star}\right\|_{2}^{2},
\end{equation}
where  %$\mathbf{x}^{\star}$ is a desired point, $i$ is the number of required iterations for RKA, and 
%$q \in \left(0,1\right)$ is 
$q=1-\frac{1}{\kappa\left(\Tilde{\bOmega}\right)}\;\in \left(0,1\right)$
\begin{comment}
\begin{equation}
\label{eq:16}
q=1-\frac{1}{\kappa\left(\Tilde{\bOmega}\right)}\;\in \left(0,1\right),
\end{equation}
\end{comment}
and $\kappa\left(\Tilde{\bOmega}\right)=\|\Tilde{\bOmega}\|^{2}_{\mathrm{F}}\|\Tilde{\bOmega}^{\dagger}\|^{2}_{2}$ is \textit{scaled condition number} \cite{edelman1992distribution} of $\Tilde{\bOmega}$, which is a block matrix of $m$ diagonal matrices per \eqref{eq:9}. We have % we first obtain $\|\Tilde{\bOmega}\|^{2}_{\mathrm{F}}$ as
\begin{equation}
\label{fro}
\|\Tilde{\bOmega}\|^{2}_{\mathrm{F}}=\sum^{m n}_{j=1}r^{2}_{j}=\sum^{m n}_{j=1}1=m n.
\end{equation}
Moreover, $\|\Tilde{\bOmega}^{\dagger}\|^{2}_{2}=\frac{1}{\sigma^{2}_{\textrm{min}}}$, where $\sigma_{\textrm{min}}=\min\left\{\sigma_{i}\right\}$ is the minimum singular value of $\Tilde{\bOmega}$ \cite{van1996matrix} (maximum singular value is $\sigma_{\textrm{max}}$ similarly defined). Following Lemma~\ref{lemma_1} evaluates singular values of $\Tilde{\bOmega}$. % is evaluated based on the following theorem to help with determining $\sigma_{\textrm{min}}$.
\begin{lemma}
\label{lemma_1}
Consider the concatenation of all $m$ sign data matrices in \eqref{eq:9}, i.e., $\Tilde{\bOmega} \in \mathbb{R}^{m n\times n}$, where $n$ is the size of the input signal and $m$ is the number of time-varying sampling thresholds. The matrix $\Tilde{\bOmega}$ is full-rank and its singular values are  
\begin{equation}
\label{singular}
\sigma_{1}=\sigma_{2}=\cdots=\sigma_{n}= \sqrt{m}.
\end{equation}
\end{lemma}
\begin{IEEEproof}
%To obtain the singular values of $\Tilde{\bOmega}$, 
Compute the square matrix
\begin{equation}
\label{proof_th}
\begin{aligned}
\mathbf{P} &= \Tilde{\bOmega}^{\top}\Tilde{\bOmega}
%&= 
%\left[\bOmega^{(1)}\vdots \bOmega^{(2)}\vdots\cdots\vdots\bOmega^{(m)}\right]
%\left[\begin{array}{c}
%\left(\bOmega^{(1)}\right)^{\top}\\
%\cdots\\
%\left(\bOmega^{(2)}\right)^{\top}\\
%\vdots\\
%\cdots\\
%\left(\bOmega^{(m)}\right)^{\top}
%\end{array}\right],\\
= \left[\begin{array}{c|c|c}
\bOmega^{(1)} &\cdots &\bOmega^{(m)}
\end{array}\right]\left[\begin{array}{c|c|c}
\bOmega^{(1)} &\cdots &\bOmega^{(m)}
\end{array}\right]^{\top}\\
&=\bOmega^{(1)}\left(\bOmega^{(1)}\right)^{\top}+\bOmega^{(2)}\left(\bOmega^{(2)}\right)^{\top}+\cdots+\bOmega^{(m)}\left(\bOmega^{(m)}\right)^{\top},\\
&= m\mathbf{I}.
\end{aligned}
\end{equation}
Hence, the eigenvalues of $\mathbf{P}$ are equal to $m$. In other words, the singular values of $\Tilde{\bOmega}$ are $\left\{\sigma_{i}\right\}_{i=1}^{n}=\sqrt{m}$.
\end{IEEEproof}
It follows from (\ref{fro}) and Lemma~\ref{lemma_1} that %, the scaled condition number is thus obtained as
%\begin{equation}
%\label{cond}
%\begin{aligned}
$\kappa\left(\Tilde{\bOmega}\right)=\frac{m n}{\sigma^{2}_{\textrm{min}}} = n$. Conventionally, the condition number of a matrix is defined as $\frac{\sigma_{\textrm{max}}}{\sigma_{\textrm{min}}}$. From Lemma~\ref{lemma_1}, all singular values are equal. Hence, $\kappa\left(\Tilde{\bOmega}\right) = n \frac{\sigma_{\textrm{max}}}{\sigma_{\textrm{min}}}$ is indeed the condition number scaled by $n$. This
%\end{aligned}
%\end{equation}
% which implies that 
leads to
\begin{align}
\label{cond}
q=\frac{n-1}{n}. 
\end{align}
Set the algorithm termination criterion to the condition
\begin{equation}
\label{eq:6565}
\mathbb{E}\left\{\left\|\mathbf{x}_{i}-\mathbf{x}^{\star}\right\|_{2}^{2}\right\}\leq \epsilon_{1},
\end{equation}
where $\epsilon_{1}$ is a positive constant. Based on this criterion and (\ref{eq:15}), the following Proposition~\ref{lemma_3} states the lower bound on the number of required RKA iterations. 
\begin{proposition}
\label{lemma_3}
The number of RKA iterations $i$ required to achieve the optimal solution $\mathbf{x}^{\star}$ of length $n$ from its one-bit samples within the error specified by (\ref{eq:6565}) is
\begin{equation}
\label{bound5}
\begin{aligned}
i&\geq \frac{\log\left(\frac{\omega_{0}}{\epsilon_{1}}\right)}{\log\left(\frac{1}{1-\frac{1}{n}}\right)},
\end{aligned}
\end{equation}
where $\omega_{0}=\left\|\mathbf{x}_{0}-\mathbf{x}^{\star}\right\|_{2}^{2}$ is the initial squared error (at $i=0$) and $\epsilon_{1}$ is a positive constant. 
\end{proposition}
\begin{IEEEproof}
%To satisfy (\ref{eq:6565}), the inequality in (\ref{eq:15}) is adjusted to capture the upper bound $\epsilon_{1}$. As a result, the following inequality is obtained:
Define
%\begin{equation}
%\label{bound31}
%\begin{aligned}
$q^{i}\left\|\mathbf{x}_{0}-\mathbf{x}^{\star}\right\|_{2}^{2}\leq \epsilon_{1}$, 
%\end{aligned}
%\end{equation}
or equivalently,
\begin{equation}
\label{bound311}
\begin{aligned}
q^{i}&\leq \frac{\epsilon_{1}}{\omega_{0}}.
\end{aligned}
\end{equation}
Note that $\omega_{0}$ is a constant scalar that depends on only the initial and optimal solutions. % and is thus considered to be fixed for the purposes of this analysis. 
Substituting
(\ref{cond}) in \eqref{bound311} and taking logarithm on both sides yields
\begin{equation}
\label{bound30}
i\log\left(1-\frac{1}{n}\right)\leq \log\left(\frac{\epsilon_{1}}{\omega_{0}}\right).
\end{equation}
%This completes the proof. 
%or equivalently,
%\begin{equation}
%\label{iiii}
%i\geq \frac{\log\left(\frac{\omega_{0}}{\epsilon_{1}}\right)}{\log\left(\frac{1}{1-\frac{1}{n}}\right)},
%\end{equation}
%which proves the theorem.
\end{IEEEproof}
Since the optimal solution is unknown, $\omega_{0}$ may not be precisely determined. However, a suitable number of required iterations may still be selected following Theorem~\ref{lemma_3} with a reasonable guess for $\omega_{0}$. For instance, an initial value $\mbx_{0}$ may be chosen in the direction of optimal solution $\mbx^{\star}$ so that a $\omega_{0}$ is obtained \cite{strohmer2009randomized,leventhal2010randomized}.

%----------------------------------------------------------------------
%\begin{figure}[t]
%	\center{\includegraphics[width=0.45\textwidth]{recovery-kaczmarz-true.eps}}
%	\caption{Reconstruction of a sawtooth wave input signal from one-bit measurements using the RKA, with the true signal plotted along its reconstructed version. %\textcolor{red}{what is the label of y-axis? why the x-axis is t? It should be sample index. Also, need to write this caption in a similar way we wrote in the ICASSP paper, with legends saying x, xbar etc. You need to also magnify portions of the plot to show how good the reconstruction is. Merely saying it is good and giving a computed NMSE is not sufficient. After all, we claim near-perfect reconstruction!} 
%	}
%	\label{figure_1}
%\end{figure}
%----------------------------------------------------------------------
%\vspace{-8pt}
\subsection{Numerical Example}
\label{NUM_RKA}
Fig.~\ref{figure_1}a illustrates the RKA reconstruction of a sawtooth signal from one-bit polyhedron in (\ref{eq:8n}) for $10$ sweeps (periods) %periods of a sawtooth wave 
with a fundamental frequency of $50~\mathrm{Hz}$. We discretized the generated signal $x(t)$ at the sampling rate (interval) of $1~\mathrm{kHz}$ ($\mathrm{T}=0.001~\mathrm{s}$). The time-varying sampling thresholds were drawn from the distribution $\boldsymbol{\uptau}^{(\ell)}\sim\mathcal{N}\left(\mathbf{0},\mathbf{I}\right)$, for all $\ell\in\mathcal{L}$. Define the normalized mean squared error, % (NMSE) as 
%\begin{equation}
%\label{eq:17}
$\operatorname{NMSE} \triangleq \frac{\left\|\mathbf{x}-\bar{\mathbf{x}}\right\|_{2}^{2}}{\left\|\mathbf{x}\right\|_{2}^{2}}$, 
%\end{equation}
where $\mathbf{x}$ and $\bar{\mathbf{x}}$ denote the true (discretized) signal and its reconstructed version, respectively. %\textcolor{red}{now, here you call xstar the true signal, not the optimal signal. } \Fr{As can be seen in Algorithm~\ref{algorithm_1}, we have already taken care of it.} \textcolor{red}{but the true value is $\mathbf{x}[\cdot]$ (discretized version of x(t). why do we need a new variable for this quantity?}
Since RKA selects each hyperplane randomly in each iteration, we repeat the reconstruction in Fig.~\ref{figure_1}a for $15$ times. %\textcolor{red}{no noise, right? then, why do you need to repeat? The max i should remain the same too.} \Fr{We should definitely run the algorithm for several times. The reason behind this stems from two facts: (1) If our input signal have some noise with specific distribution, then averaging the NMSE is required to show the efficacy of our proposed algorithm (in this example, we did not include any noise in our signal, thus, this fact is not the reason behind the averaging here). (2) Our proposed algorithm is the RKA, which selects the $j$-th hyperplane, \emph{randomly}, for obtaining the ($i+1$)-th solution $\mathbf{x}_{i+1}$ from the solution of previous iteration $\mathbf{x}_{i}$. The convergence of the RKA was proved in the expectation manner, therefore, one cannot necessarily expect to have the same number of iterations for each solution path (different experiments). If one can obtain the convergence of the RKA without the expectation manner, then that would be a great contribution, however, it's still an open problem.} \textcolor{red}{ok. explain this briefly here why we need to run so many times and look at the average}
The averaged NMSE over all experiments is only $\sim 0.0012$ or  $-29.2082~\mathrm{dB}$. % where its value is averaged over $15$ experiments.
%----------------------------------------------------------------------
%\begin{figure}[t]
%	\center{\includegraphics[width=0.45\textwidth]{recovery-kaczmarz-false.eps}}
%	\caption{Reconstruction of a \textcolor{blue}{high DR} \textcolor{red}{just provide the DR values for Figs. 1, 2, and 3 than using relative/informal terms like `ultra-high', `super high' etc. You can use `extremely high'} input signal from one-bit measurements using the RKA, with the true signal plotted along its reconstructed version. \textcolor{red}{what is the label of y-axis? why the x-axis is t? It should be sample index. Also, need to write this caption in a similar way we wrote in the ICASSP paper, with legends saying x, xbar etc.} \Fr{maybe you can replot them and add legends and everything in the format? Because I will use MATLAB for the plot and it will be different from our ICASSP.}}
%	\label{figure_2}
%\end{figure}
%----------------------------------------------------------------------
%\vspace{-8pt}
\subsection{Limitations of Conventional One-Bit Reconstruction}
%\subsection{Reconstruction Challenges In Conventional One-Bit Sampling}
\label{Guaranteed}
%This section is devoted to the challenges one would face in the reconstruction of the one-bit sampled signal by deploying RKA due to a dynamic range mismatch. Denote the dynamic range of a signal $\mathbf{x}$ as $\text{DR}_{\mathbf{x}}=\|\mathbf{x}\|_{\infty}$. 
Denote the DRs of the desired signal $\mathbf{x}$ and the time-varying threshold $\boldsymbol{\uptau}$ by $\text{DR}_{\mathbf{x}}$ and $\text{DR}_{\boldsymbol{\uptau}}$, respectively, where we define the DR of a vector as its $\ell_{\infty}$-norm. \emph{%If $\text{DR}_{\mathbf{x}} \leq \text{DR}_{\mathbf{\uptau}}$, then we may have a chance to reconstruct the signal of interest $\mathbf{x}^{\star}$ within the polyhedron (\ref{eq:8n}) with a high probability depending on the number of samples. 
If $\text{DR}_{\mathbf{x}} \leq \text{DR}_{\boldsymbol{\uptau}}$, then the reconstructed signal $\mathbf{x}^{\star}$ may be found inside the polyhedron (\ref{eq:8n}) with a high probability for an \textup{adequate} number of samples. Otherwise, if $\text{DR}_{\mathbf{x}} > \text{DR}_{\boldsymbol{\uptau}}$, there is no guarantee to obtain $\mathbf{x}^{\star}$ since the desired solution cannot be inside the finite-volume space imposed by the set of inequalities in (\ref{eq:8n}) indicating an irretrievable information loss.} We demonstrate this as follows. Without loss of generality, consider $x_{k}=\text{DR}_{\mathbf{x}}$ for $x_{k}>0$. Assume $\tau_{k}^{\star}=\max_{\ell}~\tau_{k}^{(\ell)}$. Since $\text{DR}_{\boldsymbol{\uptau}}=\|\boldsymbol{\uptau}\|_{\infty}$, we have $\tau_{k}^{\star}\leq\text{DR}_{\boldsymbol{\uptau}}$. If $\text{DR}_{\mathbf{x}} > \text{DR}_{\boldsymbol{\uptau}}$, then we have $\tau_{k}^{\star}<\text{DR}_{\mathbf{x}}=x_{k}$. Therefore, to reconstruct the $k$-th entry of the input signal $x_{k}$, we always have a gap $\delta=x_{k}-\tau_{k}^{\star}>0$ that is not covered by any sample to capture the amplitude information of $\mathbf{x}$. Hence, the desired signal is not found inside the finite-volume space imposed by the %set of 
inequalities in (\ref{eq:8n}).

%Consider the example presented in Section~\ref{NUM_RKA}. 
In Fig.~\ref{figure_1}a, %dynamic range (DR) \textcolor{red}{already defined DR in the intro. That is where dynamic range was used first} of the time-varying sampling threshold is 
$\text{DR}_{\boldsymbol{\uptau}}=3$ %\textcolor{blue}{with a high probability} \textcolor{red}{what do you mean? we are choosing the threshold. so, why high probability here?\Ae{since the thresholds generated based on the proposition~2, is a Gaussian process with $\sigma_{\boldsymbol{\uptau}}=\lambda/3$ so its DR with high probability (0.99) is $\lambda$.} Do you mean expected/average value?}\Ae{No! I mean the probability, in this case high probability means 0.99 pls see proposition~2} \textcolor{red}{but Prop 2 comes later. right now, when you say high probability, the reader does not know why. so you need to add some explanation here} 
is larger than %the dynamic range of the input signal 
$\text{DR}_{\mathbf{x}}=1$ thereby leading to a low reconstruction NMSE. % performance as shown in Fig.~\ref{figure_1}. 
We now consider $x$ to be a bandlimited function with piece-wise constant Fourier transform %\textcolor{blue}{Fourier spectrum}. \textcolor{red}{CTFT or DTFT? or just the `spectrum'?} \Fr{just spectrum} The signal \Fr{spectrum} \textcolor{red}{is it the signal or its spectrum?} \Fr{the spectrum}\textcolor{red}{then why you say signal values? spectrum is either CTFT or DFT. specify which one is it here?}
values are drawn uniformly at random, i.e., $\widehat{x}(\omega)\sim\operatorname{unif}(0,1)$.  This signal is the same as the one used in \cite{bhandari2020unlimited}. %\textcolor{blue}{We generated $15$ realizations.} \textcolor{red}{you do not mention the NMSE. So, where were these 15 realizations even used? } \Fr{To verify Fig. 3, we have generated the signal $15$ times. We cannot make a conclusion only using one observation} 
The time-varying sampling thresholds were generated following the procedure explained in Section~\ref{NUM_RKA}. Fig.~\ref{figure_1}b shows the RKA-based reconstruction of the bandlimited signal from the polyhedron (\ref{eq:8n}). Around $t=0$ (corresponding sample index is $364$ in the plot), the reconstruction severely degrades because $\text{DR}_{\mathbf{x}}=5$ is set to be larger than $\text{DR}_{\boldsymbol{\uptau}}=3$. Indeed, when the difference between $\text{DR}_{\mathbf{x}}$ and $\text{DR}_{\boldsymbol{\uptau}}$ increases further, we observe a significant loss of information in the reconstructed signal (Fig.~\ref{figure_1}c). 
%----------------------------------------------------------------------
%\begin{figure}[t]
%	\center{\includegraphics[width=0.45\textwidth]{recovery-kaczmarz-false-two.eps}}
%	\caption{Reconstruction of an ultra-high DR input signal from one-bit measurements using the RKA, with the true signal plotted along its reconstructed version. \textcolor{red}{same comment as in the other two figures. Also, this figure and Fig. 2 should be subfigures of the same figure because they illustrate the same concepts. Even Fig. 1 could be a subfigure in the same figure. \Fr{yes I agree that all Fig~1 to 3 can be presented in one figure} When you have different figures that show the same quantities but differ only in some variable, than do not repeat the entire caption for each figure. It is not a good writing practice Instead, use the style `As in ...' See, for example, captions of Figs. 4, 5, and 6 in my paper \protect\url{https://arxiv.org/pdf/2208.04381.pdf} } }
%	\label{figure_3}
%\end{figure}
%----------------------------------------------------------------------
%\vspace{-8pt}
\section{Toward a Reconstruction Guarantee for One-Bit Sampling}
\label{sec:uno}
Since RKA does not guarantee an exact signal reconstruction from one-bit measurements in (\ref{eq:8n}) when the DR of the signal exceeds that of the time-varying sampling threshold, %This is typically the mismatch between the dynamic ranges of the input signal and the time-varying sampling threshold.  
it is pertinent to design the time-varying sampling threshold such that $\text{DR}_{\mathbf{x}}\leq\text{DR}_{\mathbf{\boldsymbol{\uptau}}}$. This is not always possible because the desired signal is unknown. We address this limitation via UNO, which is our proposed new one-bit sampling method based on the concept of unlimited sampling. %An overview of the unlimited sampling is presented in Section~\ref{sec:unlimited}. Then, our proposed sampling structure, \emph{UN}limited \emph{O}ne-bit (UNO) sampling, will be discussed in detail in Section~\ref{psm}. Section~\ref{numerical_unlim} is dedicated to numerically evaluate the efficacy of our approach in signal reconstruction. Finally, the error reconstruction for UNO algorithm (devised signal reconstruction within the UNO sampling framework) will be investigated in a probabilistic sense in Section~\ref{error_UNO}.
%------------------------------------------------
%\begin{figure}[t]
%	\center{\includegraphics[width=0.48\textwidth]{didid.eps}}
%	\caption{The UNO sampling architecture. \textcolor{red}{what is being multiplied to in the multiplier? Don't you need to multiply two quantities?}\Ae{Here, I meant the delta function we have in sampling the input signal.I did it same as paper Unlimited sampling please see their paper.} \textcolor{red}{yes. but when you draw a multiplier, you need two quantities to be precise. We need not copy A.B.'s diagram. If we can explain better, then we use the above diagram. Just answer my questions and fill in the ?? in the labels} }
%	\label{figure_30}
%\end{figure}
%------------------------------------------------

%\subsection{Unlimited Sampling Meets One-Bit Quantization: A Roadmap to Judicious Sampling}
%\label{psm}
%Based on our discussion in Section~\ref{Guaranteed}, one cannot guarantee an exact signal reconstruction from one-bit data for high dynamic-range signals using the RKA. The reason behind this is typically the mismatch between the dynamic ranges of the input signal and the time-varying sampling threshold. To address this issue and guarantee the reconstruction process, one may wish to design the time-varying sampling threshold in such a way that $\text{DR}_{\mathbf{x}}\leq\text{DR}_{\mathbf{\uptau}}$. However, this is not always possible since the desired signal is unknown. 
As discussed in Section~\ref{sec:unlimited}, unlimited sampling yields signal amplitudes folded within the range $\left[-\lambda,\lambda\right]$. % instead of directly using point-wise samples of bandlimited function $x(t)$. 
This %observation may provide an alternative roadmap to design the
suggests an alternative time-varying threshold with the same DR as the modulo samples $\Tilde{\mathbf{x}}=[\Tilde{x}_{k}]$; i.e. $\text{DR}_{\boldsymbol{\uptau}}=\lambda$. %It appears the self-reset ADC can be integrated with one-bit sampling while guiding the generation of the time-varying thresholds in a meaningful way. 
In other words, the thresholds are modified to be closer to the clipping value and the self-reset ADC is integrated with one-bit sampling. % Therefore, by exploiting the unlimited sampling alongside the RKA, we will have a fundamentally enhanced odds at reconstructing the desired signal from the polyhedron (\ref{eq:8n}) in various dynamic range settings. 
We summarize this UNO sampling framework as follows:
\begin{enumerate}
    \item Apply the modulo operator defined in (\ref{eq:18}) to the input signal $\mathbf{x}$ and obtain modulo samples $\Tilde{\mathbf{x}}=\mathcal{M}_{\lambda}\left(\mathbf{x}\right)$.
    \item Design sequences of the time-varying sampling threshold $\left\{\boldsymbol{\uptau}^{(\ell)}\right\}_{\ell=1}^{m}$ such that $\left|\text{DR}_{\boldsymbol{\uptau}^{(\ell)}}-\lambda\right|\leq \varepsilon_{0}$  
    for all $\ell\in\mathcal{L}=\{1,\cdots,m\}$ and a small number $\varepsilon_{0}$.
    \item Apply the one-bit quantization to modulo samples as $\mathbf{r}^{(\ell)}=\operatorname{sgn}\left(\Tilde{\mathbf{x}}-\boldsymbol{\uptau}^{(\ell)}\right)$.
\end{enumerate}

\begin{figure}[t]
	\centering
	\includegraphics[width=1.0\textwidth]{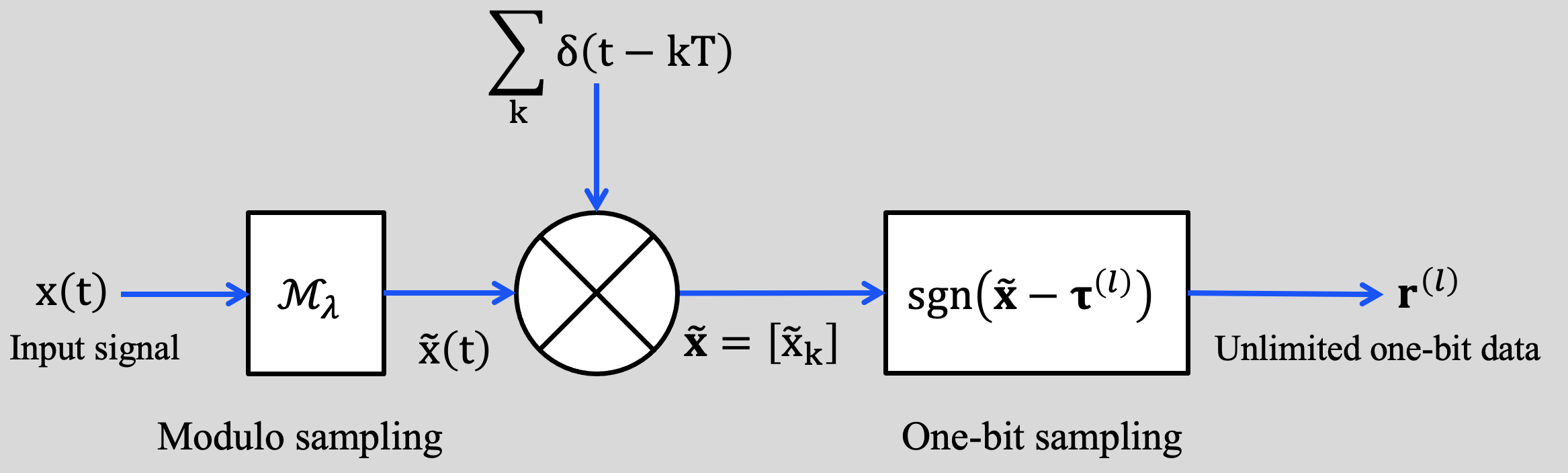}
	\caption{The UNO sampling architecture. The proper choice of the sampling interval $T$ in the middle block is specified by Theorem~\ref{Negar_Theorem}. %Note that ADC is included in the one-bit sampling block. 
    %\vspace{-10pt}
	}
	\label{figure_30}
\end{figure}

Fig.~\ref{figure_30} illustrates various steps of our UNO sampling technique. The following Proposition~\ref{UNO_prop} states the UNO threshold design.
\begin{proposition}[Judicious threshold design]
\label{UNO_prop}
Under the UNO sampling framework, the following \textup{DR guarantee} holds: Assume each one-bit sampling threshold $\boldsymbol{\uptau}^{(\ell)}$ is distributed as $\boldsymbol{\uptau}^{(\ell)}\sim \mathcal{N}\left(\mathbf{0},\sigma^{2}_{\boldsymbol{\uptau}}\mathbf{I}\right)$. Then, considering the ADC threshold $\lambda$, $\sigma_{\boldsymbol{\uptau}}$ will be equal to $\frac{\lambda}{3}$ with a probability of at least $0.99$.
\end{proposition}
\begin{IEEEproof}
With a probability of at least $0.99$, the DR of each $\boldsymbol{\uptau}^{(\ell)}\sim \mathcal{N}\left(\mathbf{0},\sigma^{2}_{\mathbf{\uptau}}\mathbf{I}\right)$ is $3\sigma_{\boldsymbol{\uptau}}$ \cite{kendall1987kendall}.
When $\sigma_{\boldsymbol{\uptau}}=\frac{\lambda}{3}$, then time-varying sampling threshold also has a DR of $\lambda$ with a probability of at least $0.99$.
\end{IEEEproof}

%------------------------------------------------
%------------------------------------------------

%------------------------------------------------
\begin{figure}[t]
	\centering
	\subfloat[Transfer function of 1-bit ADC]
		{\includegraphics[width=0.48\textwidth]{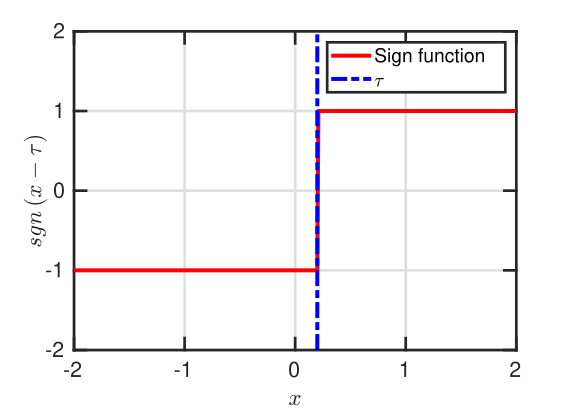}}\quad
	\subfloat[Input signal and threshold]
		{\includegraphics[width=0.48\textwidth]{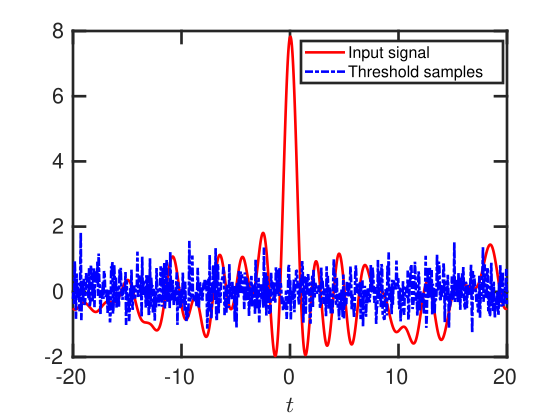}}
		\qquad
	\subfloat[Transfer function of UNO ADC]
		{\includegraphics[width=0.48\textwidth]{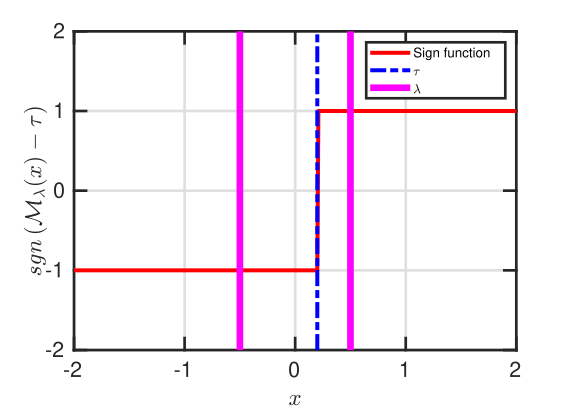}}\quad
	\subfloat[Modulo samples and threshold]
		{\includegraphics[width=0.48\textwidth]{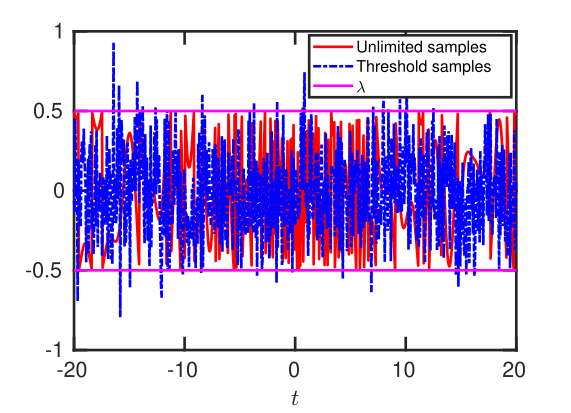}}
	\caption{%Comparison between two sampling techniques: 
	(a) Transfer function of conventional one-bit ADC where the $i$-th element of the input signal $x=(\mathbf{x})_i$ is compared with a randomly selected threshold $\boldsymbol{\uptau}$ (b) High DR input signal $\mathbf{x}$ and its thresholds samples $\uptau$. (c) As in (a), but for UNO with the judicious time-varying threshold $\lambda$. (d) The unlimited samples $\Tilde{\mbx}$ compared with the thresholded samples $\boldsymbol{\uptau}$ and $\lambda$.
	%\vspace{-12pt}
	}
	\label{figure_7nm}
\end{figure}
%-------------------------------------------------

%------------------------------------------------
\begin{figure*}[t]
	\centering
	\subfloat[$m=2$]
		{\includegraphics[width=0.33\textwidth]{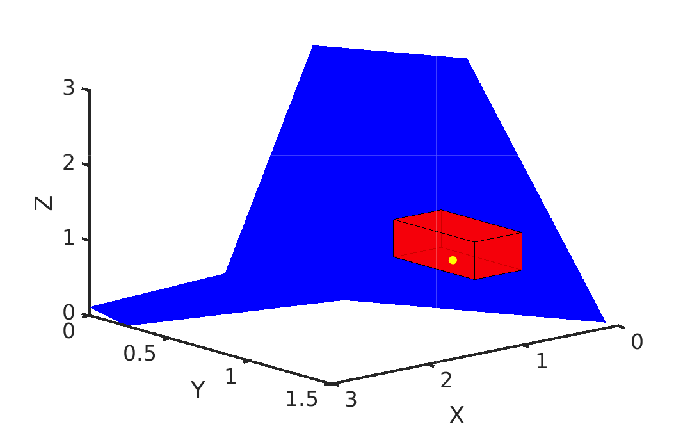}}
	\subfloat[$m=6$]
		{\includegraphics[width=0.33\textwidth]{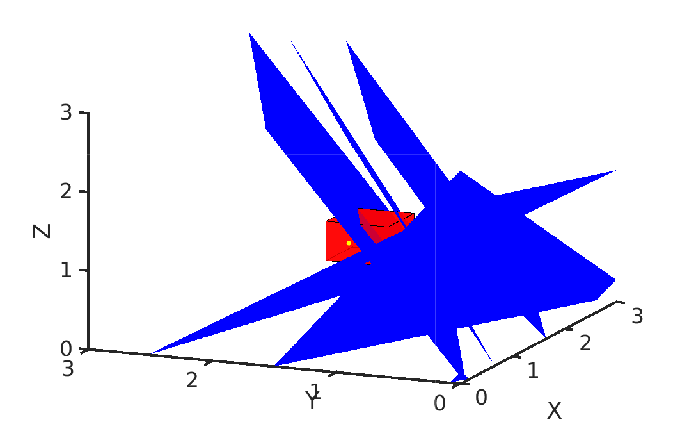}}
	\subfloat[$m=20$]
		{\includegraphics[width=0.33\textwidth]{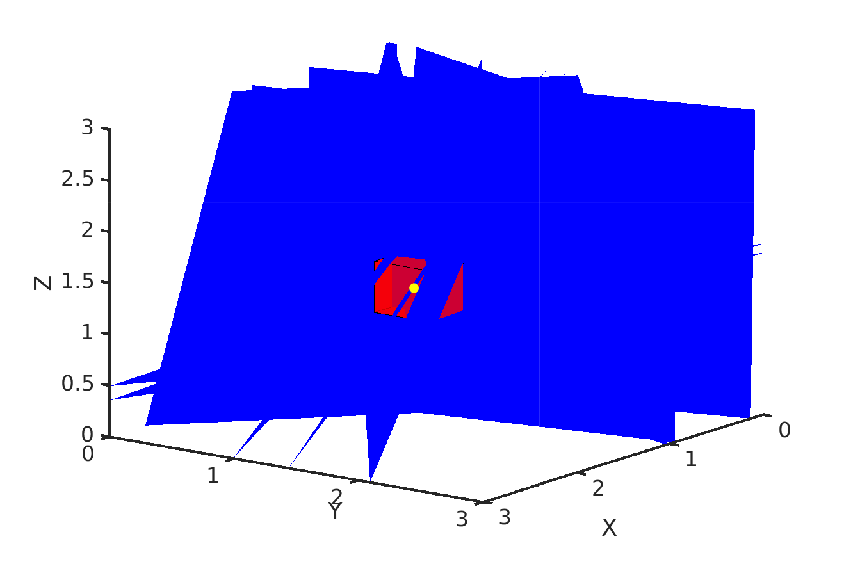}}
	\qquad	
	\subfloat[$m=2$]
		{\includegraphics[width=0.33\textwidth]{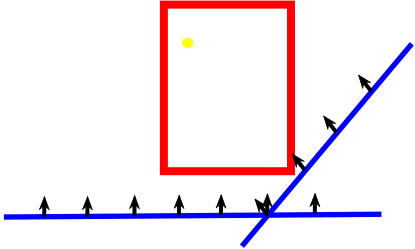}}
	\subfloat[$m=6$]
		{\includegraphics[width=0.33\textwidth]{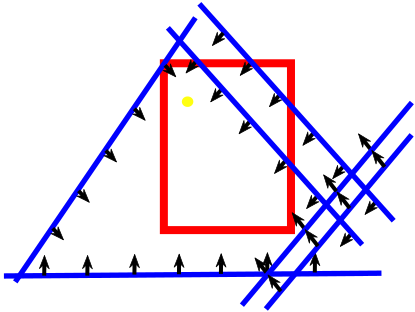}}
	\subfloat[$m=20$]
		{\includegraphics[width=0.33\textwidth]{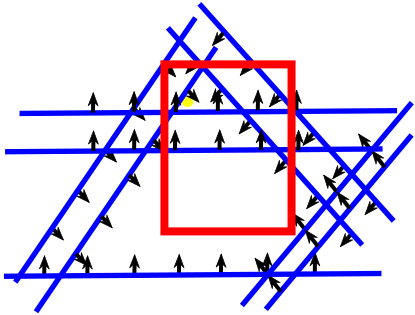}}
	
	\caption{Top: Trihedron space (polyhedron (\ref{eq:24}) in $3$ dimensions) (blue), unlimited sampling cube (red), and true value of the modulo signal $\Tilde{\mbx}\in\mathbb{R}^{3}$ (yellow) for (a) $m=2$ (b) $m=6$ and (c) $m=20$. Bottom: As in the top panel, but only a cross-section (unshaded with same color boundary) at $Z=0$ plane is shown for (d) $m=2$ (e) $m=6$ and (f) $m=20$.
	%shown with contours and its red boundary, when the number of constraints (samples) grows large. 
	Each inequality constraint is shown by a half-space whose feasible region %with  \textcolor{red}{main side??\Ae{yes, each half space has its own main side!}} 
 is marked by black arrows. %\textcolor{red}{unclear what this means. rephrase} %The evolution of the feasible regime with increasing samples is depicted in three cases: (a) and (d) small sample-size regime, constraints not forming a finite-value polyhedron; (b) and (e) medium sample-size regime, constraints forming a finite-volume polyhedron, parts of which are outside the cube; (c) and (f) large sample-size regime, constraints forming a finite-volume polyhedron inside the cube, making the signal reconstruction achievable. The optimal point representing the signal to be reconstructed is shown by yellow. VIJAY: already mentioned all these details in the text. No need to repeat again and make the caption extremely long. Instead, clearly describe the plot and what is being plotted
    %\vspace{-10pt}
	}
	\label{figure_1n}
\end{figure*}
%------------------------------------------------

In Proposition~\ref{UNO_prop}, we design time-varying sampling threshold sequences so that their DR is close to that of the input signal. This enables storing the information of distance between the input signal and the thresholds without any loss of information via one-bit sampling. Fig.~\ref{figure_7nm} shows a comparison of conventional one-bit sampling and UNO for the high DR scenario; the transfer function of the former is plotted in Fig.~\ref{figure_7nm}a. We consider the same bandlimited signal as in Section~\ref{Guaranteed} 
%To show the potential of UNO in comparison to conventional one-bit sampling, we visually display the difficulties of the one-bit sampling in the high-dynamic range regimen in Fig.~\ref{figure_7nm}. As can be seen in a-b, when we compare the input signal generated same as Section~\ref{Guaranteed}, 
and a random threshold $\boldsymbol{\uptau}\sim\mathcal{N}\left(\mathbf{0},\mathbf{I}\right)$. In case of one-bit sampling, the signal values and thresholds differ considerably at some points %the thresholded samples are indifferent to the dynamic  range \textcolor{blue}{of the input} 
(Fig.~\ref{figure_7nm}b) and, consequently, the information on the distance between the signal value and the threshold samples is completely lost. For UNO, the threshold is chosen closer to the folded signal with $\lambda=0.5$ (Fig.~\ref{figure_7nm}c). This preserves the information of the input signal in the modulo samples (Fig.~\ref{figure_7nm}d). %unlimited samples for the one-bit quantization. 
%Note that  the thresholding is most informative when the threshold is close to the range of the signal. Vijay: already said this last line in the previous para

For reconstruction of the signal of interest $\mathbf{x}$ %\textcolor{red}{same issue. what is the signal of interest? True signal x, reconstructed signal hatx or barx. you said earlier xstar is a true signal. But then it is also an optimal signal at other places!}\Ae{it is true signal value as you mentioned earlier} \Fr{yes, exactly. Signal of interest is always the true signal.}\textcolor{red}{but the true value is $\mathbf{x}[\cdot]$ (discretized version of x(t). why do we need a new variable for this quantity?} \Ae{Ok, change all xstar to x except in the convergence analysis.}
from UNO samples, we
reformulate the polyhedron (\ref{eq:8n}) for modulo samples as
 \begin{equation}
 \label{eq:24}
   \Tilde{\mathcal{P}} = \left\{\Tilde{\mathbf{x}} \mid \Tilde{\bOmega} \Tilde{\mathbf{x}} \succeq \operatorname{vec}\left(\mathbf{R}\right)\odot \operatorname{vec}\left(\bGamma\right)\right\}.
\end{equation}
This overdetermined system of linear inequalities in (\ref{eq:24}) is then solved via RKA and, from the resulting approximated modulo samples, we obtain $\mathbf{x}$ %\textcolor{red}{but above you just said that you want to get xstar!}\Ae{changed all $\mathbf{x}^{\star}$ to true value} \textcolor{red}{but the true value is $\mathbf{x}[\cdot]$ (discretized version of x(t). why do we need a new variable for this quantity?}
via Algorithm ~\ref{algorithm_1}.
%proceed as follows:
%\begin{enumerate} \textcolor{red}{you already enumerate all of these in Algorithm 1, so no need to create another list here. Just explain it in a para}
%    \item Reformulate the polyhedron (\ref{eq:8n}) for modulo samples as
%    \begin{equation}
%    \label{eq:24}
%   \Tilde{\mathcal{P}} = \left\{\Tilde{\mathbf{x}} \mid \Tilde{\bOmega} \Tilde{\mathbf{x}} \succeq \operatorname{vec}\left(\mathbf{R}\right)\odot \operatorname{vec}\left(\bGamma\right)\right\}.
%    \end{equation}
%    \item Solve the overdetermined linear system of inequalities in (\ref{eq:24}) using RKA.
%    \item Obtain the desired signal $\mathbf{x}$ from approximated modulo samples using the unlimited sampling reconstruction procedure presented in Algorithm~\ref{algorithm_1}.
%\end{enumerate}
Algorithm~\ref{algorithm_2} summarizes these steps of  %Algorithm~\ref{algorithm_2}, 
the \emph{UNO algorithm}. 
\clearpage
%-------------------------------------------------
\begin{algorithm}[H] 
	\caption{%UNO-based input 
	Signal reconstruction in UNO.}
    \label{algorithm_2}
    \begin{algorithmic}[1]
    \Statex \textbf{Input:} Sequences of one-bit measurements $\left\{\mathbf{r}^{(\ell)}=\operatorname{sgn}\left(\mathcal{M}_{\lambda}\left(\mathbf{x}\right)-\boldsymbol{\uptau}^{(\ell)}\right)\right\}_{\ell=1}^{m}$, $\boldsymbol{\uptau}^{(\ell)}\sim \mathcal{N}\left(\mathbf{0},\sigma^{2}_{\boldsymbol{\uptau}}\mathbf{I}\right)$, ADC threshold $\lambda$, total number of iterations $i_{\textrm{max}}$. 
    \Statex \textbf{Output:}  The approximation of the input signal $\bar{\mathbf{x}}$. %\textcolor{red}{reconstructed signal uses hat, which you also use for Fourier transform}
    \vspace{1.2mm}
    %\textcolor{red}{are steps 1-3 sequential or simulatneous? If it is the latter, then write all of them in the same line.}
    \State %Compute 
    $\mbR \gets \left\{\mbr^{(\ell)}\right\}_{\ell=1}^{m}$.
    \vspace{1.2mm}
    \State $\sigma_{\boldsymbol{\uptau}} \gets \frac{\lambda}{3}$
     \vspace{1.2mm}
    \State  $\bGamma \gets \left\{\boldsymbol{\uptau}^{(\ell)}\right\}_{\ell=1}^{m}$
    \vspace{1.2mm}
    \State $\bOmega^{(\ell)} \gets \operatorname{diag}\left(\mathbf{r}^{(\ell)}\right)$
    \vspace{1.2mm}
   % \textcolor{red}{But you do not use $\bOmega$ anywhere in the subsequent steps }
    \State $\Tilde{\bOmega}\gets\left[\begin{array}{c|c|c}
\bOmega^{(1)} &\cdots &\bOmega^{(m)}
\end{array}\right]^{\top}$ 
    \vspace{1.2mm}
   % \textcolor{red}{I do not see where you got $\bOmega^{(m)}$ etc. in the previous steps } 
    \State
    $\Tilde{\mathcal{P}}\gets \left\{\Tilde{\mathbf{x}} \mid \Tilde{\bOmega} \Tilde{\mathbf{x}}\succeq \operatorname{vec}\left(\mathbf{R}\right)\odot \operatorname{vec}\left(\bGamma\right)\right\}$.
    \vspace{1.2mm}
    \State Find the modulo signal in $\Tilde{\mathcal{P}}$ via RKA.
    
    \For{$i=1:i_{\textrm{max}}$} %\textcolor{red}{In Algorithm 1, you use subscript with parentheses for iteration-related variables. Here, you do not.}

    \State $\Tilde{\mathbf{x}}_{i+1}\gets\Tilde{\mathbf{x}}_{i}+\left(-\upomega_{j}\Tilde{\mathbf{x}}_{i}+\upomega_{j}\boldsymbol{\uptau}^{(\ell)}\right)^{+}\upomega^{\top}_{j}$ %$\triangleright$ $\upomega_{j}$ is the $j$-th row of $\Tilde{\bOmega}$. 
    \EndFor
    \State $\bar{\Tilde{\mbx}}\gets \Tilde{\mathbf{x}}_{i_{\textrm{max}}}$ %\textcolor{red}{and now you have both bar and hat OMG! Also, a calligraphic subscript!!}
    \State Reconstruct the input signal via Algorithm~\ref{algorithm_1} from $\bar{\Tilde{\mbx}}$. %\textcolor{red}{return command missing}
    \State \Return $\bar{\mbx}$
    %\textcolor{red}{return is the last step. algorithm ends at return}
    \end{algorithmic}
\end{algorithm}
%-------------------------------------------------

In Fig.~\ref{figure_1n}, we show %numerical investigation of 
that increasing the number $m$ of time-varying sampling threshold sequences guarantees the RKA-based reconstruction as it leads the space formed by the intersection of half-spaces (inequality constraints in \eqref{eq:24}) to completely shrink to the true value modulo signal $\Tilde{\mbx}$ inside the volume space imposed by unlimited sampling. This volume space is a \emph{cube} because the constraints applied to the modulo samples are $-\lambda\leq\Tilde{x}_{k}\leq\lambda$. Here, the blue planes/lines representing the linear inequalities form a finite-volume space around the optimal point (displayed by the yellow circle 
inside the cube) by increasing the number of one-bit sampling thresholds. In the top panel, we show the specific case of a trihedron (i.e., modulo samples are $\Tilde{\mbx}\in\mathbb{R}^{3}$) to represent the effect of increasing the number of threshold sequences on the reconstruction performance. The bottom panel shows the same effect for 2-D cross-section of the trihedron. The constraints are not enough to create a finite-volume space in Fig.~\ref{figure_1n}a and d. On the other hand, in Fig.~\ref{figure_1n}b and e, such constraints create the desired finite-volume polyhedron space but are unable to capture the optimal point. Finally, in Fig.~\ref{figure_1n}c and f, the optimal point is successfully captured by the resulting finite-volume space. The following theorem summarizes the UNO guarantees.

%------------------------------------------------
\begin{figure*}[t]
	\centering
	\includegraphics[width=1.0\textwidth]{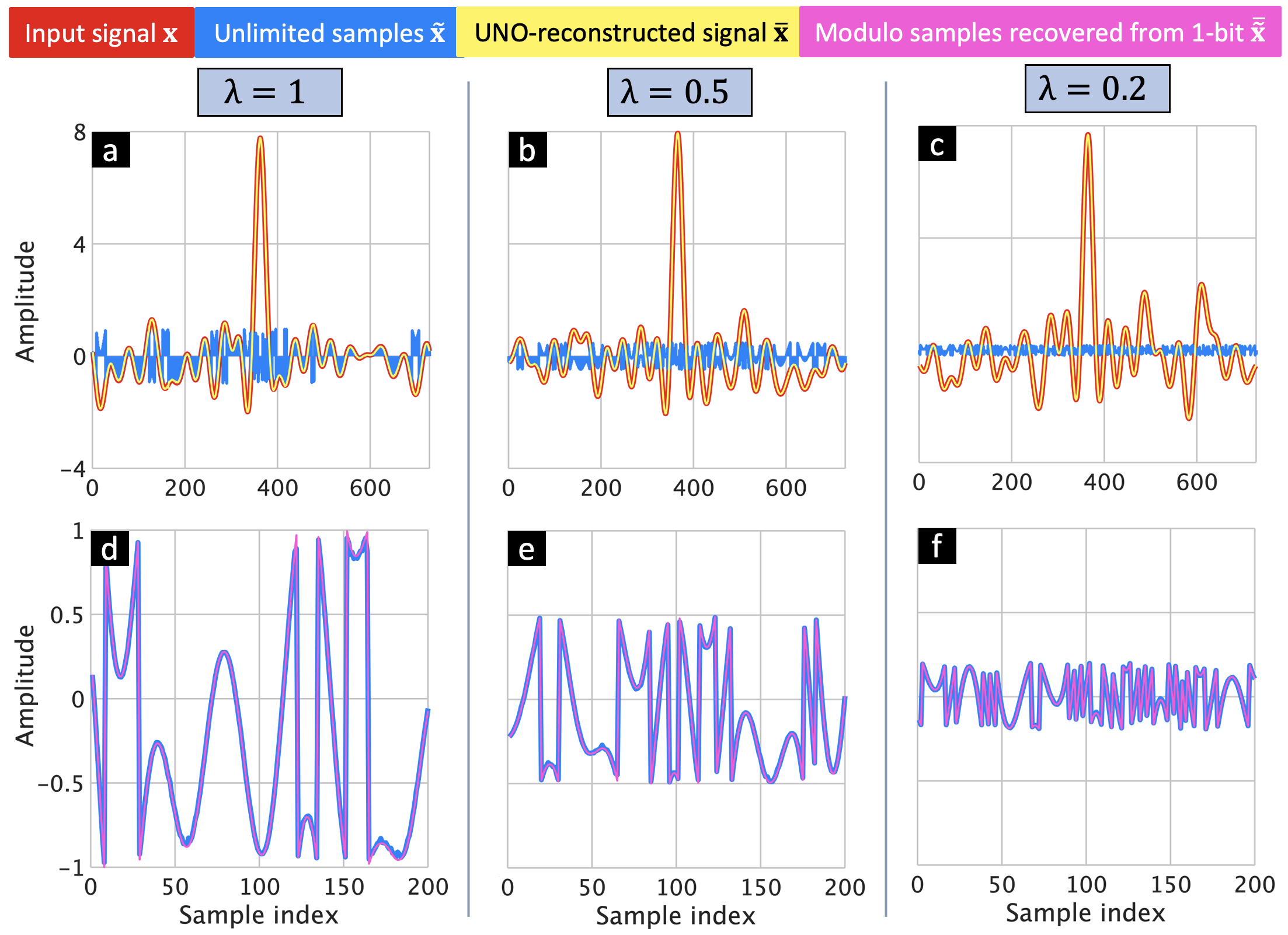}
	%\subfloat[$\lambda=1$]
	%	{\includegraphics[width=0.32\textwidth]{unlimited-L-1.eps}}
	%\subfloat[$\lambda=0.5$]
	%	{\includegraphics[width=0.32\textwidth]{unlimited-L-0.5.eps}}
	%\subfloat[$\lambda=0.2$]
	%	{\includegraphics[width=0.32\textwidth]{unlimited-L-0.2.eps}}
	%\qquad	
	%\subfloat[$\lambda=1$]
	%	{\includegraphics[width=0.32\textwidth]{modular-unlimited-L-1.eps}}
	%\subfloat[$\lambda=0.5$]
	%	{\includegraphics[width=0.32\textwidth]{modular-unlimited-L-0.5.eps}}
	%\subfloat[$\lambda=0.2$]
	%	{\includegraphics[width=0.32\textwidth]{modular-unlimited-L-0.2.eps}}
		
	\caption{Reconstruction of the input signal from one-bit measurements using UNO when the ADC threshold is (a) $\lambda=1$, (b) $\lambda=0.5$, and (c) $\lambda=0.2$. (d)-(f) As in, respectively, (a)-(c) but the true unlimited samples are compared with their reconstructed samples. %signal reconstruction within the polyhedron (\ref{eq:24}) using the RKA is investigated by  in (d)-(f). 
    %\vspace{-10pt}
	}
	\label{figure_4}
\end{figure*}
%------------------------------------------------

\begin{theorem}[UNO sampling theorem]
\label{Negar_Theorem}
Assume $x(t)$ to be a finite energy, bandlimited signal with maximum frequency $\Omega_{\textrm{max}}$. Let $\Tilde{x}_{k}$, $k \in \mathbb{Z}$, introduced in (\ref{eq:18}) be the modulo samples of $x(t)$ with sampling rate $1/T$. Assume $\bar{\Tilde{\mbx}}$ contains the modulo samples reconstructed by the RKA and define the reconstruction error as $\mbe=\left(\bar{\Tilde{\mbx}}-\Tilde{\mbx}\right)$. Then, the sufficient condition for the reconstruction of bandlimited samples $x_{k}$ from UNO samples $\left\{\mathbf{r}^{(\ell)}=\operatorname{sgn}\left(\mathcal{M}_{\lambda}\left(\mathbf{x}\right)-\boldsymbol{\uptau}^{(\ell)}\right)\right\}_{\ell=1}^{m}$, where $\boldsymbol{\uptau}^{(\ell)}\sim \mathcal{N}\left(\mathbf{0}, \frac{\lambda^{2}}{9}\mathbf{I}\right)$, up to additive multiples of $2\lambda$ is
\begin{equation}
\label{rageN}
T \leq \frac{1}{2^{h}\Omega_{\textrm{max}} e},
\end{equation}
where $h\in\mathbb{N}$ is given by
\begin{equation}
\label{rageN1}  
h\geq \frac{\log\left(\frac{2\beta_{x}}{\lambda}\right)}{\log\left(\frac{\lambda}{4\left\|\mbe\right\|_{\infty}}\right)},
\end{equation}
and
\begin{equation}
\label{rageN2}
\lambda\geq 4 \zeta\left\|\mbe\right\|_{\infty}, \quad \zeta>1.
\end{equation}
\end{theorem}
\begin{IEEEproof}
\label{Negarianproof}
While reconstructing the modulo samples from one-bit data via RKA, the real modulo samples are represented by the linear model
\begin{equation}
\label{rageN3}   
\bar{\Tilde{\mbx}}=\Tilde{\mbx}+\mbe.
\end{equation}
The error in RKA reconstruction may be viewed as \emph{noise} for modulo samples. According to \cite[Theorem 3]{bhandari2021unlimited}, the sampling rate for the \emph{contaminated} modulo samples in \eqref{rageN3} to reconstruct the bandlimited samples $x_{k}$ to satisfy $\bar{x}_{k}=x_{k}+e_{k}$ is $T \leq \frac{1}{2^{h}\Omega_{\textrm{max}} e}$, where $h\in\mathbb{N}$, and 
\begin{equation}
\label{rageN4}
\left\|\mbe\right\|_{\infty}\leq\frac{\lambda}{4} \left(\frac{2\beta_{x}}{\lambda}\right)^{-\frac{1}{h}}.
\end{equation}
Clearly, \eqref{rageN1} follows from \eqref{rageN4}. Moreover, to ensure that the log function used in \eqref{rageN1} is positive, we have $\frac{\lambda}{4\left\|\mbe\right\|_{\infty}}\geq\zeta>1$ leading to a lower bound for the ADC threshold $\lambda\geq 4\zeta\left\|\mbe\right\|_{\infty}$. This completes the proof.
\end{IEEEproof}
Theorem~\ref{Negar_Theorem} provides %to guarantee the perfect signal reconstruction from UNO samples, \eqref{rageN} must be met. Moreover, 
the lower bound for the ADC threshold $\lambda$ in Eq.~(\ref{rageN2}). The upper bound on $T$ for UNO sampling is lower than or equal to that of the unlimited sampling (the equality holds when $h=1$) which associates with a higher sampling rate in UNO. As mentioned later in Section~\ref{subsec:sig_amp}, oversampling is a a common scenario in one-bit quantization techniques and not a major concern in UNO implementation. Note that the resulting error $\mbe$ of RKA is different than the noise considered in \cite[Theorem 3]{bhandari2021unlimited} in the sense that, unlike the latter, the corresponding reconstructed modulo samples in UNO obey $\left|\bar{\Tilde{x}}_{k}\right|<\lambda$. This ensures that $N$ in (\ref{eq:21}) guarantees $\Delta^{N}\bar{\mbx} \equiv \mathcal{M}_\lambda\left(\Delta^{N} \bar{\mbx}\right)$ or equivalently $\Delta^{N}\bar{\mbx}\equiv \mathcal{M}_{\lambda}\left(\Delta^{N}\bar{\Tilde{\mbx}}\right)$; we refer the reader to \cite{bhandari2020unlimited} for more details on this aspect. As a result, %unlike \cite[Theorem 3]{bhandari2021unlimited}, in the light of Theorem~\ref{Negar_Theorem} 
UNO perfectly reconstructs the input samples $x_{k}$ in the sense that $\bar{x}_{k}=x_{k}+e_{k}$ (up to additive multiples of $2\lambda$) with the same $N$ considered in the noiseless unlimited sampling reconstruction of \cite[Section IV.B]{bhandari2021unlimited}.

\section{UNO Reconstruction: Numerical Illustrations and Error Analyses}
\label{numerical_unlim}
We assessed the performance of the UNO reconstruction through extensive numerical experiments. %algorithm in two ways: analyzing both the impact of (i) using different values for the ADC threshold $\lambda$, and (ii) using various input signal amplitudes in the reconstruction results.
In particular, we validate that the size of the cube imposed by self-reset ADCs (red contours and shaded regions in Fig.~\ref{figure_1n}) and, hence, the reconstruction error depend on the ADC threshold $\lambda$. We then investigate the effect of input signal amplitude $\left\|\mbx\right\|_{\infty}$ on the reconstruction performance. In all experiments, we considered the same high DR input signal as in Section~\ref{Guaranteed}. 

\begin{figure*}[t]
	\centering
	\includegraphics[width=1.0\textwidth]{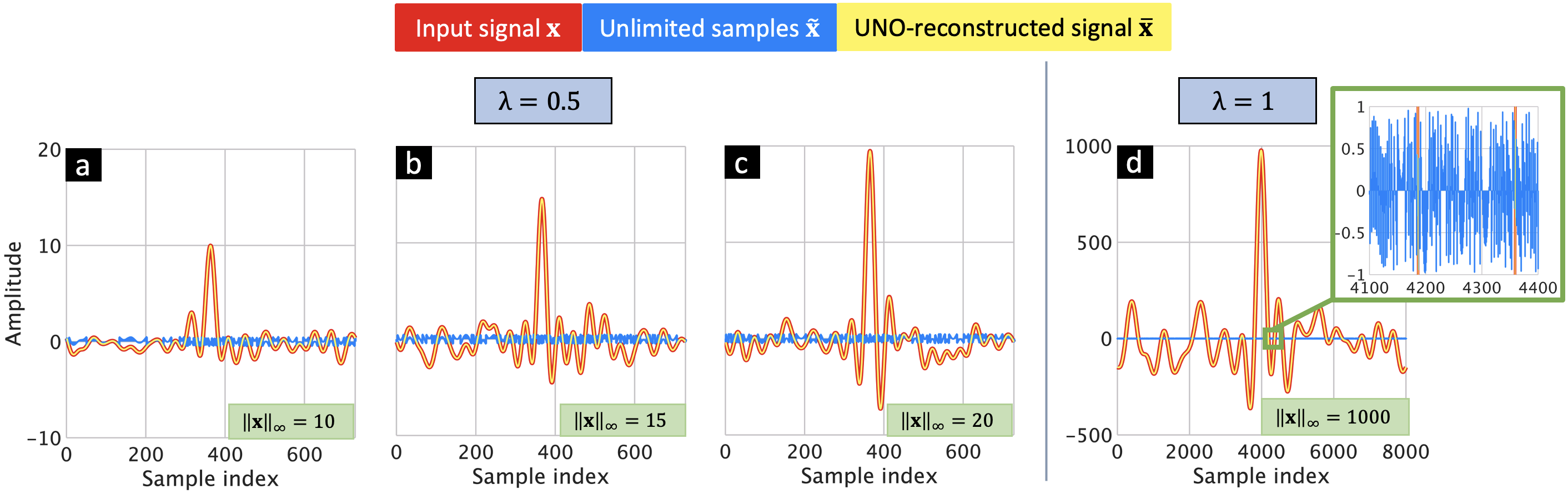}
	%\subfloat[$\left\|\mbx\right\|_{\infty}=10$]
	%	{\includegraphics[width=0.32\textwidth]{unlimited-amp-10.eps}}
	%\subfloat[$\left\|\mbx\right\|_{\infty}=15$]
	%	{\includegraphics[width=0.32\textwidth]{unlimited-amp-15.eps}}
	%\subfloat[$\left\|\mbx\right\|_{\infty}=20$]
	%	{\includegraphics[width=0.32\textwidth]{unlimited-amp-20.eps}}
	\caption{Reconstruction of the input signal from one-bit measurements using UNO Algorithm~\ref{algorithm_2} when the ADC threshold is set to $\lambda=0.5$ and the input signal amplitude $\left\|\mbx\right\|_{\infty}$ is (a) $10$, (b) $15$, and (c) $20$. (d) As in (a) but for $\lambda=1$ and $\left\|\mbx\right\|_{\infty}=1000$. The inset shows the same plot on a larger scale.
    %\vspace{-10pt}
    }
	\label{figure_5}
\end{figure*}
%\vspace{-8pt}
\subsection{Varying ADC Threshold}% $\blambda$} 
The number of time-varying sampling thresholds was set to $m=400$. In each experiment, the generated signals have the same $\mathrm{DR}_{\mbx}=8$ but the ADC threshold $\lambda$ changes. For a given $\lambda$, the sequences of time-varying sampling threshold are drawn randomly following the distribution $\left\{\boldsymbol{\uptau}^{(\ell)}\sim \mathcal{N}\left(\mathbf{0},\frac{\lambda^{2}}{9}\mathbf{I}\right)\right\}_{\ell=1}^{m}$. Fig.~\ref{figure_4} illustrates accurate UNO reconstruction for different values of $\lambda \in \{0.2,0.5,1\}$. %As can be seen, the UNO algorithm appears to guarantee the reconstruction of the input signal from one-bit measurements via solving the overdetermined linear system of inequalities (\ref{eq:24}) using the RKA in the case of different values for the ADC threshold $\lambda$. 
Table~\ref{table_1} lists the reconstruction NMSE (on a $\log_{10}$ scale), averaged over $15$ experiments, for different values of $\lambda$. % associated with the reconstructed signal $\hat{\mathbf{x}}$ in Fig.~\ref{figure_4}, where the results are averaged over $15$ experiments.
We observe that increasing in $\lambda$ leads to higher NMSE because the volume of the unlimited sampling cube grows further, and consequently, more hyperplanes may be required to contain a specific volume around the optimal point in the feasible region.
%\vspace{-8pt}
\subsection{Varying Input Signal Amplitude} 
\label{subsec:sig_amp}
 Here, we generated the input signals %similar to Section~\ref{Guaranteed} by taking
 with varying DRs. %different dynamic ranges into account. 
 In each experiment, the ADC threshold $\lambda$ was fixed to $\lambda=0.5$, for which %For this specific $\lambda$, 
 we generated sequences of time-varying sampling threshold as $\left\{\boldsymbol{\uptau}^{(\ell)}\sim \mathcal{N}\left(\mathbf{0},\frac{1}{36}\mathbf{I}\right)\right\}_{\ell=1}^{m}$. Fig.~\ref{figure_5} shows accurate UNO reconstruction for different values of $\left\|\mbx\right\|_{\infty}$. %while the input signal amplitude is changing in each experiment. It can be inferred from Fig.~\ref{figure_5} that our proposed algorithm can guarantee the reconstruction performance even in the case of different input signal dynamic ranges. 
 Table~\ref{table_2} reports the corresponding NMSE %associated with Fig.~\ref{figure_5} 
 averaged over $15$ experiments.

Next, we study the reconstruction for a signal with
an extremely high DR, with %we generate the input signal and threshold sequences same as before at which
$\left\|x(t)\right\|_{\infty}=1000$. In theory, the unlimited sampling theorem guarantees reconstruction with $T \leq \frac{1}{2\Omega_{\textrm{max}} e}$. However, in practice, signal reconstruction from unlimited samples has its own limitations due to error propagation by the finite-difference operator. % at the heart of the reconstruction process. 
Specifically, for a large DR of input signal compared to that of the ADC threshold $\lambda$, the order of difference operator $N$ should also be large. But a large $N$ would also amplify the quantization/round-off noise, leading to an unstable reconstruction. In this scenario, more samples (given by the oversampling factor) are required to decrease $N$. Note that, unlike conventional ADCs, an abundant number of samples does not lead to an increase in power consumption, manufacturing cost, and per-bit chip area in one-bit ADCs. Fig.~\ref{figure_5}d shows an accurate UNO reconstruction for $\lambda=1$ and a 40 times higher sampling rate $1/T$ than the previous experiments. %As can be seen, our proposed algorithm can guarantee the reconstruction performance in the case of extremely high dynamic range input signal.

%------------------------------------------
\begin{table} [t]
\caption{Averaged UNO reconstruction NMSE for fixed $\mathbf{x}$ 
%NMSE results associated with the performance of Algorithm~\ref{algorithm_2} in input signal reconstruction when the ADC threshold $\lambda$ is changing in each experiment.
}
\centering
\begin{tabular}{  c || c | c | c }
\hline
\textbf { $\lambda$ } & $0.2$ & $0.5$ & $1$ \\ 
\hline 
\textbf { $10\log_{10}\operatorname{NMSE}$ %($\log_{10}$ scale) 
}  & $-72.721$ & $-67.660$ & $-60.987$ \\
\hline
\end{tabular}
%\vspace{-10pt}
\label{table_1}
\end{table}
%------------------------------------------

\begin{figure}[t]
	\center{\includegraphics[width=0.55\textwidth]{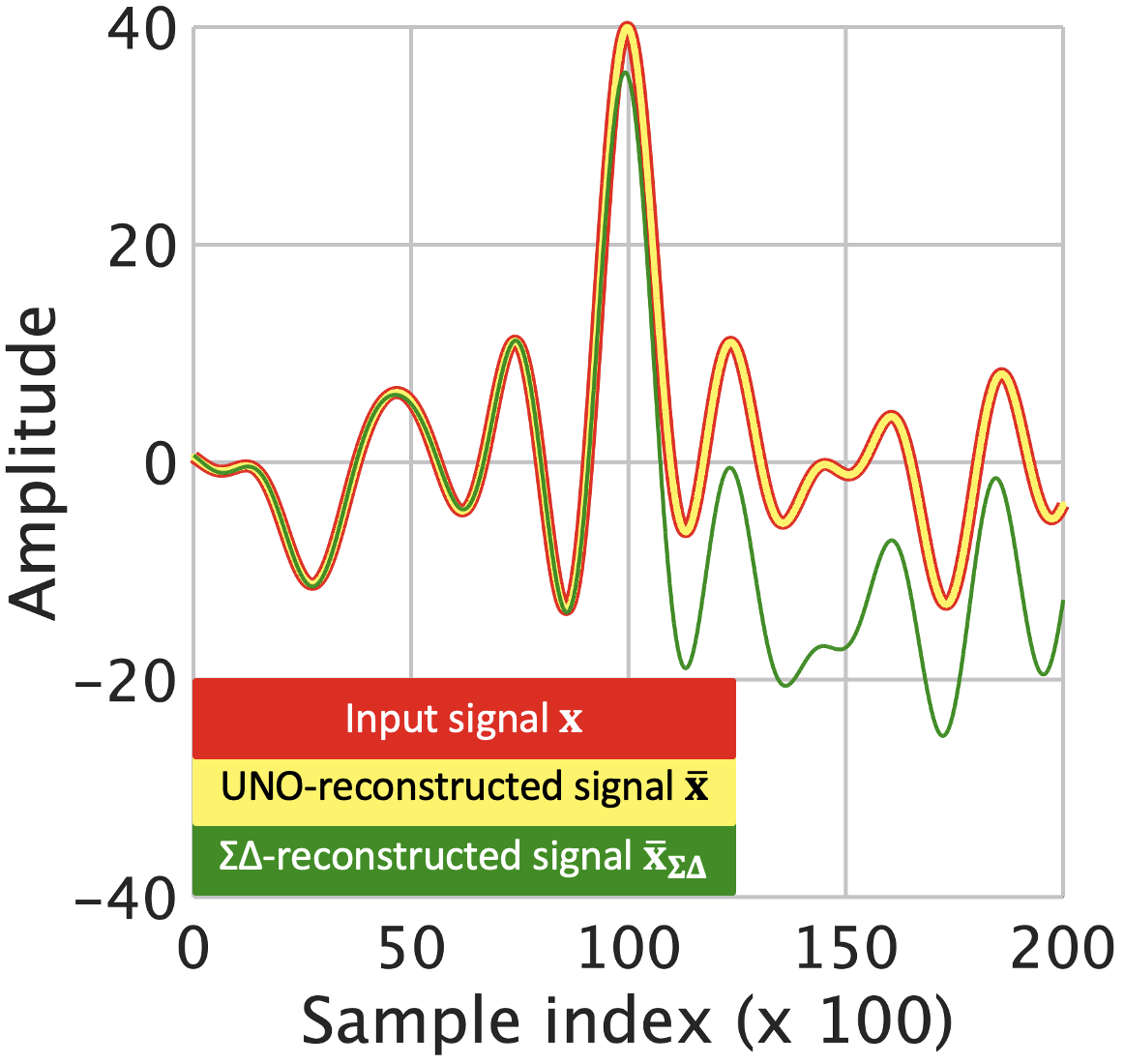}}
	\caption{A comparison of reconstruction via UNO and one-bit unlimited $\Sigma\Delta$ when $\lambda=1$ and $\left\|\mbx\right\|_{\infty}=40$. 
    %\vspace{-10pt}
    }
	\label{figure_6m}
\end{figure}

Although UNO and one-bit $\Sigma\Delta$ method \cite{graf2019one} are different in their respective theoretical foundations and applications, here we compare their reconstruction performance for the same signal. %UNO reconstruction with that of one-bit unlimited $\Sigma\Delta$ method of \cite{graf2019one}. Accordingly, 
The ADC threshold was set to $\lambda=1$ and sequences of the time-varying sampling threshold were drawn as $\left\{\boldsymbol{\uptau}^{(\ell)}\sim \mathcal{N}\left(\mathbf{0},\frac{1}{9}\mathbf{I}\right)\right\}_{\ell=1}^{m}$. For the specific case of $\left\|\mbx\right\|_{\infty}=40$, Fig.~\ref{figure_6m} compares the UNO-reconstructed signal $\bar{
\mathbf{x}}$ with the one-bit unlimited $\Sigma\Delta$-reconstructed signal $\bar{\mbx}_{\Sigma\Delta}$ when the ratio between the input signal amplitude and the ADC threshold $\eta=\frac{\left\|\mbx\right\|_{\infty}}{\lambda}$ is large. % high $\eta$ input signals, the input signal is generated with the setting same as Subsection~\ref{Guaranteed} with $\left\|\mbx\right\|_{\infty}=40$. As can be observed in Fig.~\ref{figure_6m}, 
The one-bit unlimited $\Sigma\Delta$ degenerates in some parts of the input samples, 
%especially when the \textcolor{blue}{main amplitude} \textcolor{red}{what do you mean by this term?} of the signal 
while the UNO accurately reconstruct the signal. Table~\ref{table_3} further compares the reconstruction NMSE, averaged over $15$ experiments, of both sampling methods for different amplitudes %Here, we generated input signals with different amplitudes
$\left\|\mbx\right\|_{\infty}\in\left\{20,50\right\}$. % input signal reconstruction accuracy of UNO and one-bit $\Sigma\Delta$ in term of $\operatorname{NMSE}$. As can be seen, 
%The UNO reconstruction outperforms one-bit $\Sigma\Delta$ for large $\eta$. %The reason behind the performance of one-bit $\Sigma\Delta$ in input signal reconstruction for a large value of $\eta$ is (i) round-off noise in software and more dominantly, due to (ii) imperfect noise shaping in sigma-delta conversion which results in samples corruption. 
Here, the degradation in one-bit $\Sigma\Delta$ reconstruction for large $\eta$ is because of the round-off noise in software and, primarily, imperfect noise shaping in sigma-delta conversion that results in sample corruption. %One can minimize the noise in sigma-delta conversion to a certain extent, however, it cannot be eliminated.
%---------------------------------------------------------
\begin{table}[t]
\caption{%NMSE results associated with the performance of Algorithm~\ref{algorithm_2} in input signal reconstruction when the input signal amplitude is changing in each experiment. 
Averaged UNO reconstruction NMSE for $\lambda=0.5$}
\centering
\begin{tabular}{  c || c | c | c   }
\hline
\textbf {$\left\|\mbx\right\|_{\infty}$} & $10$ & $15$ & $20$ \\[0.5 ex]
\hline
\textbf {$10\log_{10}\operatorname{NMSE}$} & $-63.925$ & $-65.820$ & $-63.969$ \\
\hline
\end{tabular}
%\vspace{-10pt}
\label{table_2}
\end{table}
%------------------------------------------
%------------------------------------------
%\begin{figure}[t]
%	\center{\includegraphics[width=0.45\textwidth]{unlimited-amp-1000.eps}}
%	\caption{Reconstruction of a extremely high DR signal from one-bit measurements using UNO algorithm, with the true signal (red color) plotted along its reconstructed version (blue) when the ADC threshold is $\lambda=1$ and the input signal amplitude is $\left\|\mbx\right\|_{\infty}=1000$.}
%	\label{ghoorak}
%\end{figure}
%------------------------------------------
%------------------------------------------
%------------------------------------------
%\begin{table} [t]
%\centering
%\caption{Comparing one-bit unlimited $\Sigma\Delta$ and UNO in the reconstruction of input signals with different amplitudes (dynamic ranges) in term of $\operatorname{NMSE}$ ($\log_{10}$ scale).}
%\centering
%\begin{tabular}{  c | c | c | c }
%\hline
%\textbf{One-bit unlimited}{$\bSigma\bDelta$} & \text {$\left\|\mbx\right\|_{\infty}=20$} & \text {$\left\|\mbx\right\|_{\infty}=50$}\\ [0.5 ex]
%\hline \hline
%$\operatorname{NMSE}$ ($\log_{10}$ scale)& $0.0402$ & $0.3777$ \\
%\hline\hline
%\textbf{UNO} & \text {$\left\|\mbx\right\|_{\infty}=20$} & \text {$\left\|\mbx\right\|_{\infty}=50$}\\
%\hline \hline
%$\operatorname{NMSE}$ ($\log_{10}$ scale)& $-6.3969$ & $-6.2081$ \\
%\hline
%\end{tabular}
%\label{table_3}
%\end{table}
%------------------------------------------
%-----------------------------------------------------------------
\begin{table} [t]
\centering
\caption{Reconstruction $10\log_{10}\operatorname{NMSE}$ for $\lambda=1$}% one-bit unlimited $\Sigma\Delta$ and UNO.}
\centering
\begin{tabular}{ c | c | c }
\hline
$\left\|\mbx\right\|_{\infty}$ & \textbf{One-bit unlimited} $\bSigma\bDelta$ & \textbf{UNO}\\ [0.5 ex]
\hline \hline
$20$ & $0.402$ & $-63.969$ \\
\hline
$50$ & $3.777$ & $-62.081$ \\
\hline
\end{tabular}
%\vspace{-10pt}
\label{table_3}
\end{table}
%-----------------------------------------------------------------
%\vspace{-8pt}
\subsection{Analysis of Reconstruction Error}% Investigation for the UNO Algorithm}
\label{error_UNO}
%An integral part of our proposed reconstruction algorithm is RKA whose reconstruction error was readily discussed in Section~\ref{ERROR_BOUND}. Nevertheless, as observed in Fig.~\ref{figure_1n}, because of deploying self-reset ADCs from unlimited sampling in UNO and creation of a finite-volume space in the feasible region (cube), the number of time-varying sampling threshold sequences $m$ appears to affect the reconstruction error. VIJAY: Already said before

To ensure a bounded reconstruction error, the feasible region in (\ref{eq:24}) cannot have an infinite volume in an asymptotic sense
%\textcolor{red}{unclear what do you mean by this phrase. do you mean a polyhedron with infinite faces? or do you mean an unbounded polyhedron?}\Ae{polyhedron with infinite faces is better phrase here.} \textcolor{red}{ok. but you also say asymptotic sense. what would `infinite in asymptotic sense' change to?} \Ae{"Infinite" here means infinite volume! in the asymptotic case (large number of hyperplanes) we will have a finite volume space located around the optimal point} \Fr{done}
when amplitude constraints are imposed by unlimited sampling. As mentioned before, by introducing more samples, it is possible to obtain %the problem forms 
a polyhedron with a bounded volume that contains the desired point. Further, as we illustrated in Fig.~\ref{figure_1n}, adding more inequality constraints to (\ref{eq:24}) leads to shrinkage of this polyhedron. % will put a downward pressure on the error between the desired and reconstructed points (each informative sample will shrink this space). 
We now prove this result, i.e., in a probabilistic sense, that increasing the number of samples leads to the reconstruction error approaching zero, and that the resulting overdetermined linear system of inequalities guarantees the convergence of RKA \cite{eamaz2022phase,leventhal2010randomized,briskman2015block}.  In other words, using an abundant number of samples (or oversampling in one-bit), the probability of creating the finite-volume space around the desired point $\Tilde{\mathbf{x}}^{\star}$ is increased.

%We investigate the convergence of the UNO algorithm through a probabilistic lens. 
Define the distance between the optimal solution $\Tilde{\mathbf{x}}^{\star}$ and the $j$-th hyperplane of (\ref{eq:24}) %\textcolor{red}{how do we know that it was j-th hyperplane that was in (33)?} \Fr{herein, we simply just defined the distance between the desired point $\Tilde{\mathbf{x}}^{\star}$ and the $j$-th hyperplane presented in (\ref{eq:24}). What is the ambiguity here?} \textcolor{red}{I do not see any j in (33). }  \Fr{In the equation below, when we are defining the distance, we formulate the $j$-th hyperplane of (\ref{eq:24}), simultaneously, $\upomega_{j}$ is the $j$-th row of $\Tilde{\bOmega}$} 
as
\begin{equation}
\label{distance}
\begin{aligned}
d_{j}\left(\Tilde{\mathbf{x}}^{\star},\boldsymbol{\uptau}^{(\ell)}\right) &= \left\|\upomega_{j}\left(\Tilde{\mathbf{x}}^{\star}-\boldsymbol{\uptau}^{(\ell)}\right)\right\|^{2}_{2},\quad j\in\left\{1,\cdots,m^{\prime}\right\},
\end{aligned}
\end{equation}
where $\upomega_{j}$ is the $j$-th row of $\Tilde{\bOmega}$. This distance %$d_{j}\left(\Tilde{\mathbf{x}}^{\star},\mathbf{\uptau}^{(\ell)}\right)$ 
is also the residual error\footnote{In a linear feasibility problem $\mbC\mbx\leq \mbb$, the \emph{residual error} is defined as $\left\|\mbC\mathbf{x}^{\star}-\mathbf{b}\right\|_{2}^{2}$ \cite{van1996matrix}.} of (\ref{eq:24}). %\textcolor{red}{another new, undefined term} \Ae{residual error is a well-known term in the literature of app math when we want to work with the error between the both sides of the equation. Simply, it is $e=\left\|\mathbf{A}\mathbf{x}^{\star}-\mathbf{b}\right\|_{2}^{2}$ for a linear equation $\mathbf{A}\mathbf{x}=\mathbf{b}$ or a linear system of inequalities $\mathbf{A}\mathbf{x}\leq\mathbf{b}$}. \textcolor{red}{say what it is mathematically for (33) like how you wrote here now. I see it is also mentioned correctly in the para before Theorem 2 below} 
Intuitively, it is easy to observe that by reducing the distances between $\Tilde{\mathbf{x}}^{\star}$ and the constraint-associated hyperplanes generally increases the possibility of capturing the optimal point. For a specific sample size $m^{\prime}=m n$, when the volume of the finite space around the optimal point is reduced, the mean $\left\{d_{j}\left(\Tilde{\mathbf{x}}^{\star},\boldsymbol{\uptau}^{(\ell)}\right)\right\}_{j=1}^{m^{\prime}}$ \cite{leventhal2010randomized}, i.e.,
%\begin{equation}
%\label{ave}
%\begin{aligned}
$T_{\text{ave}}=\frac{1}{m^{\prime}}\sum^{m^{\prime}}_{j=1}d_{j}\left(\Tilde{\mathbf{x}}^{\star},\boldsymbol{\uptau}^{(\ell)}\right)$, 
%\end{aligned}
%\end{equation}
also decreases. Denote $D_{\ell}\left(\Tilde{\mathbf{x}}^{\star},\boldsymbol{\uptau}^{(\ell)}\right)=\left\|\Tilde{\mathbf{x}}^{\star}-\boldsymbol{\uptau}^{(\ell)}\right\|^{2}_{2}$. Then, $T_{\text{ave}}$ %(\ref{ave}) 
becomes %\textcolor{red}{why you do not average over m below?}\Ae{Actually, we took average. Since the sum over $d_{j}$ is equal to that of $D_{\ell}$. Note that $\upomega$ only keep one element and make others zero, so we write it vector wise in the (36)}
\begin{equation}
\label{6566}
\begin{aligned}
T_{\text{ave}} = \frac{1}{m^{\prime}}\sum^{m}_{\ell=1}D_{\ell}\left(\Tilde{\mathbf{x}}^{\star},\boldsymbol{\uptau}^{(\ell)}\right).
\end{aligned}
\end{equation}

In the one-bit phase retrieval approach studied in \cite{eamaz2022phase}, a \emph{Chernoff bound} was derived to quantify the possibility of creating the above-mentioned finite-volume and the number of samples required for RKA. We apply this result below in Theorem~\ref{theorem_0} to the UNO reconstruction from one-bit samples. Here, we have replaced the error between the true value signal and the initial value $\left\|\Tilde{\mbx}_{0}-\Tilde{\mbx}^{\star}\right\|^{2}_{2}$ with the residual error $\left\|\upomega_{j}\left(\Tilde{\mathbf{x}}^{\star}-\boldsymbol{\uptau}^{(\ell)}\right)\right\|^{2}_{2}$ in the Chernoff bound. The latter explains the distance between the hyperplanes and the true value $\mbx^{\star}$ by including the sampling threshold sequences into its expression.
\begin{theorem}\cite[Theorem 1]{eamaz2022phase}
\label{theorem_0}
Assume the distances $\left\{d_{j}\left(\Tilde{\mathbf{x}}^{\star},\boldsymbol{\uptau}^{(\ell)}\right)\right\}_{j=1}^{m}$ between the desired point $\Tilde{\mathbf{x}}^{\star}$ and the hyperplanes of the polyhedron defined in (\ref{eq:24}) are independent and identically distributed random variables. Then,
\begin{enumerate}
    \item The Chernoff bound of $T_{\text{ave}}$ is 
\begin{equation}
\label{eq:theorem_cher}
\operatorname{Pr}\left(\frac{1}{m^{\prime}}\sum^{m^{\prime}}_{j=1}d_{j}\left(\Tilde{\mathbf{x}}^{\star},\boldsymbol{\uptau}^{(\ell)}\right)\leq a\right)\geq 1-\inf_{t\geq 0}\frac{M_{T}}{e^{t a}},
\end{equation}
where $a$ is an average distance point in space at which the finite-volume space around the desired signal is created, and
\begin{equation}
\label{eq:psi}
M_{T} = \left(1+t\frac{\mu^{(1)}_{d_{j}}}{m^{\prime}}+\cdots+t^{\kappa}\frac{\mu^{(\kappa)}_{d_{j}}}{\kappa!m^{\prime\kappa}}+\mathcal{O}\left(m^{\prime}\right)\right)^{m^{\prime}},
\end{equation}
%$M_{T}$ 
is the moment generating function (MGF) of the reconstruction error, $\mu^{(\kappa)}_{d_{j}}=\mathbb{E}\left\{d^{\kappa}_{j}\right\}$, and $\mathcal{O}\left(m^{\prime}\right)$ denotes the higher-order terms. % associated with truncating the Taylor series expansion of $M_{T}$.
\item $M_{T}$ decreases with an increase in the sample size %in the sample abundance scenario, 
leading to an increase in the lower bound in (\ref{eq:theorem_cher}).
\end{enumerate}
\end{theorem}
Theorem~\ref{theorem_0} states that the abundant number of samples in conventional one-bit quantization significantly affect the reconstruction performance of RKA for a system of linear inequalities in (\ref{eq:24}). Based on this result, Claim~\ref{clm} shows the efficacy of UNO sampling.

\begin{claim}
\label{clm}
Increasing the number of time-varying sampling threshold sequences $m$ is not an effective approach to guarantee the desired signal reconstruction with RKA without using unlimited sampling.
\end{claim}
\begin{IEEEproof}
For RKA-based reconstruction in Section~\ref{sec:onebit_rec}, assume that we increase the number of time-varying sampling threshold sequences from $m$ to $m+\kappa$. Therefore, from (\ref{eq:theorem_cher}) of Theorem~\ref{theorem_0}, the Chernoff bound of the reconstruction error is 
\begin{equation}
\label{cor_1}
\operatorname{Pr}\left(\frac{1}{(m+\kappa)n}\sum^{(m+\kappa)n}_{j=1}d_{j}\left(\mathbf{x}^{\star},\boldsymbol{\uptau}^{(\ell)}\right)\leq a\right)\geq 1-P_{T},
\end{equation}
where $P_{T}=\inf_{t\geq 0}\frac{M_{T}}{e^{ta}}$ and $d_{j}\left(\mathbf{x}^{\star},\boldsymbol{\uptau}^{(\ell)}\right)=\left\|\upomega_{j}\left(\mathbf{x}^{\star}-\boldsymbol{\uptau}^{(\ell)}\right)\right\|^{2}_{2}$. Without loss of generality, assume $x_{k}=\text{DR}_{\mathbf{x}}$ for $x_{k}>0$, and $\delta=x_{k}-\text{DR}_{\boldsymbol{\uptau}}$ when $\text{DR}_{\mathbf{x}}>\text{DR}_{\boldsymbol{\uptau}}$. The infimum of the distance $\frac{1}{(m+\kappa)n}\sum^{(m+\kappa)n}_{j=1}d_{j}\left(\mathbf{x}^{\star},\boldsymbol{\uptau}^{(\ell)}\right)$ in (\ref{cor_1}) occurs when $\tau_{k}^{(\ell)}=\text{DR}_{\boldsymbol{\uptau}}$ for $\ell\in\left\{1,\cdots,m+\kappa\right\}$. As a result, this infimum is
\begin{equation}
\label{cor_2}
\begin{aligned}
\bar{d}&=\inf\left(\frac{1}{(m+\kappa)n}\sum^{(m+\kappa)n}_{j=1}d_{j}\left(\mathbf{x}^{\star},\boldsymbol{\uptau}^{(\ell)}\right)\right),\\&=\inf\left(\frac{1}{(m+\kappa)n}\sum^{(m+\kappa)(n-1)}_{j=1}d_{j}\left(\mathbf{x}^{\star},\boldsymbol{\uptau}^{(\ell)}\right)\right)+\frac{(m+\kappa)\delta}{(m+\kappa)n},\\&=\inf\left(\frac{1}{(m+\kappa)n}\sum^{(m+\kappa)(n-1)}_{j=1}d_{j}\left(\mathbf{x}^{\star},\boldsymbol{\uptau}^{(\ell)}\right)\right)+\frac{\delta}{n}.
\end{aligned}
\end{equation}
The term $\frac{\delta}{n}$ in (\ref{cor_2}) does not depend on the number of time-varying sampling thresholds $m+\kappa$. In other words, increasing the value of $m$ does not 
guarantee the reconstruction of the desired signal in the polyhedron (\ref{eq:8n}) via the RKA. This phenomenon is also observed in connection to $P_{T}$. A considerable difference between the signal values and thresholds leads to larger values of $\left\{d_{j}\left(\mathbf{x}^{\star},\boldsymbol{\uptau}^{(\ell)}\right)\right\}$ thereby increasing the moments $\left\{\mu^{(\kappa)}_{d_{j}}\right\}$ or MGF $M_{T}$. Therefore, the dependence of $P_{T}$ on $M_{T}$ is unaffected when $m$ is increased.
%However, by employing unlimited sampling and imposing amplitude constraints, distances are bounded and $T_{\text{ave}}$ (\ref{6566}) is guaranteed to be lower than a small $a$ where the volume of the created finite-space is lower than that of the cube imposed by unlimited sampling. VIJAY: This commented text is not part of the proof. So, i moved this after the proof.
\end{IEEEproof}

\begin{figure}[t]
	\centering
	\includegraphics[width=0.70\textwidth]{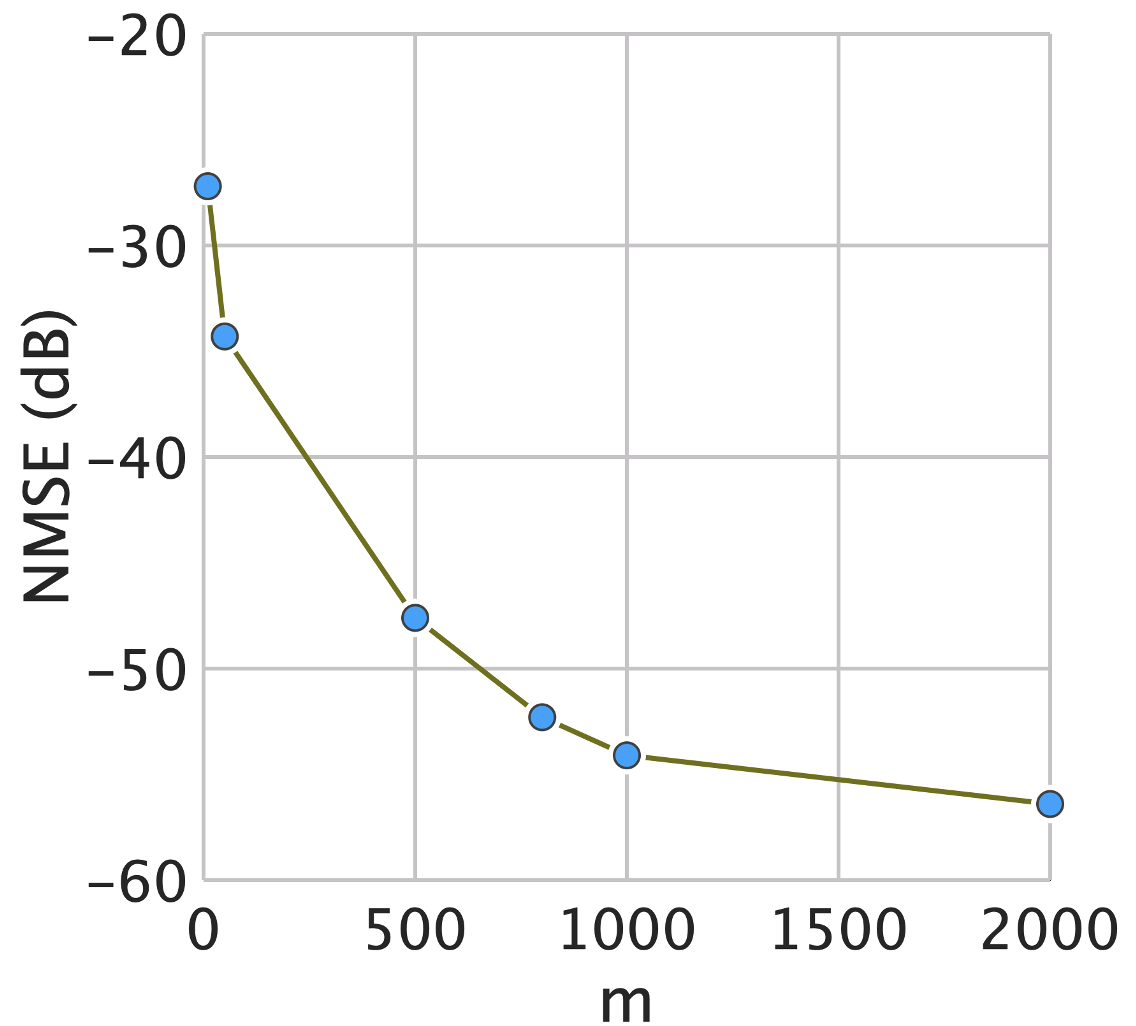}
	\caption{Average NMSE for RKA-based UNO reconstruction with respect to the number of time-varying threshold sequences $m$ for $\lambda=0.5$ and $\left\|\mbx\right\|_{\infty}=20$. 
    %\vspace{-10pt}
	}
	\label{figure_2n}
\end{figure}

Note that by using the unlimited sampling and imposing amplitude constraints, the considered distances become bounded and $T_{\text{ave}}$ (\ref{6566}) is guaranteed to be lower than a small $a$. Then, the volume of the resulting finite-space will be smaller than that of the cube imposed by unlimited sampling.
%In contrast to the above discussion, with inclusion of unlimited sampling, increasing the number of thresholds can significantly enhance the reconstruction performance. VIJAY: already said this many, many times!  
In Fig.~\ref{figure_2n}, we show that UNO reconstruction NMSE, averaged over $15$ experiments, significantly improves with the increase in the number of time-varying threshold sequences $m$. %The presented results are averaged over $15$ experiments. 
The ADC threshold was set to $\lambda=0.5$ and the signal DR was $\left\|\mbx\right\|_{\infty}=20$. 
%The number of time-varying sampling sequences utilized in UNO algorithm are $m\in \left\{50, 100, 500, 800, 1000, 2000\right\}$.
%---------------------------------------------------------
%---------------------------------------------------------

%\Ae{We investigate the relation between the number of time-varying sampling thresholds sequences and the UNO sampling rate $1/T$ in the following remark:}
\begin{remark}
 \label{rageNremark}   
According to Theorem~\ref{theorem_0}, %the reconstruction of modulo samples improves 
when the number of time-varying threshold sequences $m$ is increased, the reconstruction error $\mbe=\left(\Tilde{\mbx}-\bar{\Tilde{\mbx}}\right)$ and $\|\mbe\|_{\infty}$ become smaller. We have a smaller lower bound on $h$ defined in \eqref{rageN1} and, consequently, a lower sampling rate based on \eqref{rageN}. In other words, a larger $m$ yields a smaller UNO oversampling factor. % in the UNO sampling scheme.
\end{remark}
%By considering Remark~\eqref{rageNremark}, we try to shed light on the relation between $m$ and $T$ in the UNO sampling at which a closed form relation may be not readily found. 

\section{Reconstruction in the Presence of Noise}
%\section{Noisy-UNO Algorithm}
\label{sec:noise}
%In this section, we extend our study to signal reconstruction from noisy one-bit data under the UNO sampling framework. In most practical applications, we must rely on noisy measurements \cite{candes2015phase,davenport20141,bhaskar20151}.

%\textcolor{red}{Is there a noisy reconstruction equivalent of unlimited sampling? If yes, reproduce the theorem here and compare.} \Ae{one-bit SD has no noisy case}  \textcolor{red}{one-bit SD does not have that but unlimited sampling has a noisy reconstruction guarantee as I mentioned on p.3? That is the closest related result and should be discussed here} \Fr{Yes, it has a noisy version but it is related to the high resolution samples. Here, we discuss UNO algorithm when we have a noisy input signal. Therefore, the available information here is only one-bit. We only observed the data, partially while the noisy version of unlimited sampling has the full information (multi-bit 

In the presence of noise, one-bit $\Sigma\Delta$ sampling currently lacks similar guarantees. In one-bit noisy models of \cite{zymnis2009compressed,khobahi2020model}, a linear measurement model with additive Gaussian noise was considered. Then, based on the MLE formulation for Gaussian likelihood function, the input signal is recovered. However, in case of non-Gaussian contamination, the MLE objective is nonconvex and the recovered solution is not unique. Moreover, MLE-based reconstruction is computationally more complex for high-dimensional signals. 

Previously, for unlimited sampling, \cite{bhandari2020unlimited} has shown recovery of noisy bandlimited samples from their modulo samples up to an unknown additive constant, where the noise is entry-wise additive to the modulo samples, i.e., $\Tilde{\mby}=\Tilde{\mbx}+\bepsilon$, and $\bepsilon$ is the noise vector. Contrary to this, we propose an approach to reconstruct unlimited one-bit sampled signal when the noise is additive to the input signal, which itself has a linear relationship with a desired parameter. This linear model for the noisy measurement $\mathbf{y}$ is
\begin{equation}
\label{eq:1000}
\begin{aligned}
\mby&=\mbx+\bepsilon,\\
\mbx &= \mbA\boldsymbol{\uptheta},\quad \mathbf{A}\in\mathbb{R}^{r\times s},
\end{aligned}
\end{equation}
where $\boldsymbol{\uptheta}$ is the desired parameter vector and the noise follows the distribution $\bepsilon\sim\mathcal{N}\left(0,\sigma^{2}_{\bepsilon} \mathbf{I}_{m}\right)$. Here, we may have $\mathbf{y} \notin[-\lambda, \lambda]$. Our goal is to estimate $\boldsymbol{\uptheta}$ from the UNO samples of noisy measurement $\mby$ obtained as %. The UNO sampling of $\mathbf{y}$ yields
\begin{equation}
\label{eq:10001}
\begin{aligned}
\mathbf{r}^{(\ell)} = \operatorname{sgn}(\mathcal{M}_{\lambda}\left(\mathbf{y}\right)-\boldsymbol{\uptau}^{(\ell)}),\quad \ell\in \mathcal{L}.
\end{aligned}
\end{equation}
Our recovery approach comprises using RKA and Algorithm 1 (with N specified by (55)) to reconstruct noisy measurements from one-bit data, and then exploiting the PnP-ADMM method to estimate the desired parameters from linear overdetermined or undetermined systems.

\subsection{PnP-ADMM-Based UNO Reconstruction}
\label{sec_LASSO}

From the UNO samples \eqref{eq:10001}, we reconstruct $\mathbf{y}$ via Algorithm~\ref{algorithm_2}. The reconstructed signal $\bar{\mby}$ 
also follows the linear model (\ref{eq:1000}).  
Therefore, we use $\bar{\mby}$ to estimate $\boldsymbol{\uptheta}$ through the regularization
\begin{comment}
LASSO is a regression analysis method that performs both variable selection and regularization in order to enhance the prediction accuracy and interpretability of the resulting statistical model \cite{ali2019generalized}. This method has been extensively utilized in signal reconstruction problems by forming the optimization problem \cite{ali2019generalized}:
\begin{equation}
\label{eq:183}
\min_{\btheta}~\|\btheta\|_{1}, \text { s.t. }\|\mathbf{\hat{y}}-\mathbf{A} \btheta\|_{2} \leq \varepsilon.
\end{equation}
The minimization problem in (\ref{eq:183}) is convex which makes it attractive from a computational viewpoint. By formulating the Lagrange multiplier for the problem in (\ref{eq:183}), we have:
\begin{equation}
\label{eq:242}
\min_{\btheta,\eta}~\|\mathbf{\hat{y}}-\mathbf{A} \btheta\|_{2}^{2} + \eta \|\btheta\|_{1}
\end{equation}
This regression method can be used in both overdetermined and underdetermined systems \cite{tibshirani1996regression}. By using LASSO, the desired parameters vector $\btheta$ may be reconstructed from (\ref{eq:242}). The input additive noise can be viewed as a modeling error considered by LASSO ($\ref{eq:242}$).
\end{comment}
%To recover the signal of interest from the noisy system (\ref{eq:1000}), the following regularized formulation can be defined:
\begin{equation}
\label{Neg1}
\widehat{\boldsymbol{\uptheta}}= \arg \min_{\boldsymbol{\uptheta}}~\|\bar{\mby}-\mathbf{A} \boldsymbol{\uptheta}\|_{2}^{2} + \eta \rho(\boldsymbol{\uptheta}),
\end{equation}
where $\rho(\boldsymbol{\uptheta})$ is the penalty term and $\eta>0$ is the real-valued regularization %design 
parameter. % that trades between loss and the regularization. 
There is a rich body of literature to select the penalty function $\rho(\cdot)$ including the $\ell_{1}$-norm \cite{tibshirani1996regression}, smoothly clipped absolute deviation (SCAD) \cite{fan2001variable}, adaptive least absolute shrinkage and selection operator (LASSO) \cite{zou2006adaptive} and the minimax-concave (MC) penalty which has a relationship with Huber functions \cite{selesnick2017sparse}. 

One of the standard approaches to solve regularized problems such as in (\ref{Neg1}) is ADMM that relies on splitting variables \cite{boyd2011distributed}. We consider%\textcolor{red}{never use s.t. Follow Boyd's recommendation of using `subject to' because `s.t.' also means `such that'. }
\begin{equation}
\label{Neg2}
\widehat{\boldsymbol{\uptheta}}= \arg \min_{\boldsymbol{\uptheta}}~\|\bar{\mby}-\mathbf{A} \boldsymbol{\uptheta}\|_{2}^{2} + \eta \rho(\upnu)~\text{subject to} ~\boldsymbol{\uptheta}=\upnu.
\end{equation}
Using the augmented Lagrangian, we reformulate problem (\ref{Neg2}) as %\textcolor{red}{Follow Boyd's notation style: min is a value, minimize is the optimization task. So, the following should be `minimize max'}
\begin{equation}
\label{Neg3}
\underset{\boldsymbol{\uptheta}, \upnu}{\textrm{minimize}} \max_{\mbp}\left\{\|\bar{\mby}-\mathbf{A} \boldsymbol{\uptheta}\|_{2}^{2}+\eta \rho(\upnu)+\mbp^{\top}(\boldsymbol{\uptheta}-\upnu)+\frac{\beta}{2}\|\boldsymbol{\uptheta}-\upnu\|^2\right\},
\end{equation}
where $\mbp$ is the dual variable and $\beta$ is a real-valued design parameter. Denote $\mbu=\frac{\mbp}{\beta}$. Then, %\textcolor{red}{Follow Boyd's notation style: min is a value, minimize is the optimization task. So, the following should be `minimize max'}
\begin{equation}
\label{Neg4}
\underset{\boldsymbol{\uptheta}, \upnu}{\textrm{minimize}} \max_{\mbu}\left\{\|\bar{\mby}-\mathbf{A} \boldsymbol{\uptheta}\|_{2}^{2}+\eta \rho(\upnu)+\frac{\beta}{2}\|\boldsymbol{\uptheta}-\upnu+\mbu\|^2-\frac{\beta}{2}\|\mbu\|^2\right\}.
\end{equation}

The ADMM tackles (\ref{Neg4}) by alternating the minimization of $\boldsymbol{\uptheta}$ and $\upnu$. The update of $\upnu$ is essentially denoising of $\boldsymbol{\uptheta}_{k}+\mbu_{k-1}$ by the regularization $\eta \rho(\upnu)$. This is the key idea behind PnP-ADMM, where the proximal projection 
\begin{equation}
 \label{Neg2000}
 \upnu_{k}=\arg\min_{\upnu}\left\{\eta\rho(\upnu)+\frac{\beta}{2}\|\upnu-\boldsymbol{\uptheta}_{k}-\mbu_{k-1}\|^2\right\}
\end{equation}
is replaced with an appropriate denoiser $\mathcal{D}(.)$. For further details on various denoisers used in PnP techniques, we refer the interested reader to \cite{venkatakrishnan2013plug}.  % Image denoising has been studied for decades, with the result that high performance methods are now readily available. Today’s state-of-the-art denoisers include those based on image-dependent filtering algorithms (e.g., BM3D \cite{dabov2007image}) or deep neural networks (e.g., TNRD \cite{chen2016trainable}, DnCNN \cite{zhang2016restoration}). Most of these denoisers are not variational in nature, i.e., they are not based on any explicit regularizer $\eta \rho(\uptheta)$ \cite{reehorst2018regularization}. Leveraging the denoising interpretation of ADMM, \cite{venkatakrishnan2013plug} proposed to replace the proximal projection step for the $\upnu$ of ADMM with a call to a sophisticated image denoiser, such as BM3D, and dubbed their approach PnP ADMM. Numerical experiments have shown that PnP-ADMM works very well in most cases \cite{reehorst2018regularization,venkatakrishnan2013plug}. 
Algorithm~\ref{algorithm_3}  summarizes the noisy UNO reconstruction procedure.
%------------------------------------------------------------------
\begin{algorithm}[H]
	\caption{Noisy UNO algorithm.}
    \label{algorithm_3}
    \begin{algorithmic}[1]
    \Statex \textbf{Input:} Sequences of one-bit measurements $\left\{\mathbf{r}^{(\ell)}=\operatorname{sgn}\left(\mathcal{M}_{\lambda}\left(\mby\right)-\boldsymbol{\uptau}^{(\ell)}\right)\right\}_{\ell=1}^{m}$, where $\mby$ follows (\ref{eq:1000}), $\boldsymbol{\uptau}^{(\ell)}\sim \mathcal{N}\left(\mathbf{0},\sigma^{2}_{\boldsymbol{\uptau}}\mathbf{I}\right)$, ADC threshold $\lambda$, design parameters $\eta$ and $\beta$,    total number of iterations $k_{\textrm{max}}$.     %\textcolor{red}{calligraphic index? see all of my comments for Algo 2. They are applicable here as well}
    \Statex \textbf{Output:}  The approximation of the parameter of interest $\widehat{\boldsymbol{\uptheta}}$ .
    \vspace{1.2mm}
    \State %Compute 
    $\mbR \gets \left\{\mbr^{(\ell)}\right\}_{\ell=1}^{m}$.
    \vspace{1.2mm}
    \State $\sigma_{\boldsymbol{\uptau}} \gets \frac{\lambda}{3}$
     \vspace{1.2mm}
    \State  $\bGamma \gets \left\{\boldsymbol{\uptau}^{(\ell)}\right\}_{\ell=1}^{m}$
    \vspace{1.2mm}
    \State $\bOmega^{(\ell)} \gets \operatorname{diag}\left(\mathbf{r}^{(\ell)}\right)$
    \vspace{1.2mm}
    \State $\Tilde{\bOmega}\gets\left[\begin{array}{c|c|c}
\bOmega^{(1)} &\cdots &\bOmega^{(m)}
\end{array}\right]^{\top}$ 
%\textcolor{red}{see my comments about Omega in Algo 2}
    \vspace{1.2mm}
    \State 
    $\Tilde{\mathcal{P}}\gets \left\{\bar{\Tilde{\mathbf{y}}} \mid \Tilde{\bOmega} \bar{\Tilde{\mathbf{y}}}\succeq \operatorname{vec}\left(\mathbf{R}\right)\odot \operatorname{vec}\left(\bGamma\right)\right\}$ $\triangleright$ $\bar{\Tilde{\mby}}$ reconstructed modulo samples from RKA.
    \vspace{1.2mm}
    \State Reconstruct $\bar{\mby}$ from $\bar{\Tilde{\mby}}$ with Algorithm~\ref{algorithm_1}. %\textcolor{red}{comments are not Algo steps. They should be Statex, not State}
    \vspace{1.2mm}
    \For{$k=1:k_{\textrm{max}}$}

    \State $\boldsymbol{\uptheta}_{k}\gets\min_{\boldsymbol{\uptheta}}\left\{\|\bar{\mby}-\mathbf{A} \boldsymbol{\uptheta}\|_{2}^{2}+\frac{\beta}{2}\|\boldsymbol{\uptheta}-\upnu_{k-1}+\mbu_{k-1}\|^2\right\}$.
\vspace{1.2mm}
    \State $\upnu_{k}\gets \mathcal{D}\left(\boldsymbol{\uptheta}_{k}+\mbu_{k-1}\right).$ 
 %\textcolor{red}{just write this step here and mention the details in the text about this specific step. This is an Algorithm chart. Can't have the step in the comment}
\vspace{1.2mm}
    \State $\mbu_{k}\gets\mbu_{k-1}+\boldsymbol{\uptheta}_{k}-\upnu_{k}$.
    
    \EndFor
    \vspace{1.2mm}
    \State \Return $\widehat{\boldsymbol{\uptheta}}\gets \boldsymbol{\uptheta}_{k_{\textrm{max}}}$. %\textcolor{red}{missing return command}
    \end{algorithmic}
\end{algorithm}
%-------------------------------------------------

%\subsection{Numerical Investigation of Noisy-UNO Algorithm}
%We investigate the effectiveness of the proposed algorithm in  \ref{sec_LASSO} when applied to noisy measurements in order to estimate the desired parameters. We begin by defining an interval for ADC threshold $\lambda$ which is to be properly selected in order to guarantee the reconstruction in the noisy scenario. Next, a numerical assessment of our proposed algorithm is presented.

\subsection{ADC Threshold Selection in Noisy UNO}
%\textcolor{red}{`Noisy UNO' in the title but `Noisy-UNO' later. May be just call it NUNO :-)}\Ae{https://www.nuno.com/en/  :)}
%We investigate the effectiveness of the proposed algorithm in  \ref{sec_LASSO} when applied to noisy measurements in order to estimate the desired parameters. We begin by defining an interval for ADC threshold $\lambda$ which is to be properly selected in order to guarantee the reconstruction in the noisy scenario. Next, a numerical assessment of our proposed algorithm is presented.
%As discussed in the Section~\ref{sec_LASSO}, to estimate the desired parameter vector $\uptheta$ in (\ref{eq:1000}), our proposed noisy-UNO algorithm consists of two stages: (1) reconstructing the noisy measurements $\mathbf{y}$ by the UNO algorithm described in Algorithm~\ref{algorithm_2}, and (2) employing PnP-ADMM formulation to estimate the desired parameter vector $\uptheta$ from reconstructed measurements $\widehat{\mathbf{y}}$. VIJAY: Already said all of this so many times before!

%The following Theorem~\ref{theorem_4} states the UNO reconstruction guarantee in the presence of noise.
Theorem~\ref{theorem_4} certifies that the additive noise to the input signal results in an additive noise in modulo domain.
%establishes a bound on the ADC threshold $\lambda$  to guarantee the reconstruction of noisy measurements $\mathbf{y}$ via Algorithm~\ref{algorithm_3}, up to an unknown additive constant.
\begin{theorem}
\label{theorem_4}
Assume the noise vector in the measurement model $\mby=\mbx+\mbz$ to be $\mathbf{z}=\left[z_{k}\right]\sim \mathcal{N}\left(0,\sigma^{2}_{\mathbf{z}} \mathbf{I}_{m}\right)$. Denote $\Tilde{\mbx}=\mathcal{M}_{\lambda}\left(\mbx\right)$ and $\Tilde{\mbz}=\left[\Tilde{z}_{k}\right]$, $\Tilde{z}_{k}=\operatorname{mod}\left(z_{k},2\lambda\right)-2(1-q_{k})\lambda$, $q_{k}\in\{0,1\}$. Then, %the presence of additive input noise $\mathbf{z}$ will result in an additive noise in modulo domain $\Tilde{\mbz}$ such that
\begin{equation}
\label{eq:n_1}
\Tilde{\mby}=\Tilde{\mbx}+\Tilde{\mbz},
\end{equation}
where $\Tilde{\mby}=\mathcal{M}_{\lambda}\left(\mby\right)$.
\end{theorem}
\begin{IEEEproof}
Applying the modulo operator $\mathcal{M}_{\lambda}$ in (\ref{eq:18}) to the noisy measurements $\mathbf{y}$ produces
\begin{equation}
\label{eq:n_2}
\begin{aligned}
\Tilde{\mathbf{y}}&=\mathcal{M}_{\lambda}\left(\mathbf{y}\right)=\mathcal{M}_{\lambda}\left(\mathbf{x}+\mathbf{z}\right)\\&=\mathbf{x}+\mathbf{z}-2\lambda\left\lfloor\frac{\mathbf{x}}{2\lambda}+\frac{1}{2}+\frac{\mathbf{z}}{2\lambda}\right\rfloor,
\end{aligned}
\end{equation}
where $\mathbf{z}\sim \mathcal{N}\left(0,\sigma^{2}_{\mathbf{z}} \mathbf{I}_{m}\right)$. Since we have $\left\lfloor a+b\right\rfloor\geq\left\lfloor a\right\rfloor+\left\lfloor b\right\rfloor$ for two arbitrary real numbers $a$ and $b$, it follows from (\ref{eq:n_2}) that
\begin{equation}
\label{eq:n_3}
\begin{aligned}
\mathcal{M}_{\lambda}\left(\mathbf{x}+\mathbf{z}\right)&=\mathbf{x}+\mathbf{z}-2\lambda\left\lfloor\frac{\mathbf{x}}{2\lambda}+\frac{1}{2}+\frac{\mathbf{z}}{2\lambda}\right\rfloor\\&\leq\mathbf{x}-2\lambda\left\lfloor\frac{\mathbf{x}}{2\lambda}+\frac{1}{2}\right\rfloor+\mathbf{z}-2\lambda\left\lfloor\frac{\mathbf{z}}{2\lambda}\right\rfloor\\&=\Tilde{\mathbf{x}}+\mathbf{z}-2\lambda\left\lfloor\frac{\mathbf{z}}{2\lambda}\right\rfloor=\Tilde{\mathbf{x}}+\operatorname{mod}\left(\mathbf{z},2\lambda\right).
\end{aligned}
\end{equation}
%Similarly, the following bound can be obtained 
Using the identity $\left\lfloor a+b\right\rfloor\leq\left\lfloor a\right\rfloor+\left\lfloor b\right\rfloor+1$, we obtain
\begin{equation}
\label{eq:n_4}
\mathcal{M}_{\lambda}\left(\mathbf{x}+\mathbf{z}\right)\geq\Tilde{\mathbf{x}}+\operatorname{mod}\left(\mathbf{z},2\lambda\right)-2\lambda.
\end{equation}
%Interestingly, $\mathcal{M}_{\lambda}\left(x_{k}+z_{k}\right)$ can be written as 
A binary combination of the right-hand sides of (\ref{eq:n_3}) and (\ref{eq:n_4}) is equivalent to $\mathcal{M}_{\lambda}\left(x_{k}+z_{k}\right)$, i.e.,
\begin{equation}
\label{morad}
\begin{aligned}
\Tilde{y}_{k}&=\mathcal{M}_{\lambda}\left(x_{k}+z_{k}\right)\\
&=q_{k}\left(\Tilde{x}_{k}+\operatorname{mod}\left(z_{k},2\lambda\right)\right)+(1-q_{k})\left(\Tilde{x}_{k}+\operatorname{mod}\left(z_{k},2\lambda\right)-2\lambda\right),\\&=\Tilde{x}_{k}+\operatorname{mod}\left(z_{k},2\lambda\right)-2(1-q_{k})\lambda,
\end{aligned}
\end{equation}
where $q_{k}\in\{0,1\}$. Rewrite (\ref{morad}) as
\begin{equation}
\label{ghoor}
\Tilde{y}_{k}=\Tilde{x}_{k}+\Tilde{z}_{k},
\end{equation}
where $\Tilde{z}_{k}=\operatorname{mod}\left(z_{k},2\lambda\right)-2(1-q_{k})\lambda$, which completes the proof.
\end{IEEEproof}
It follows from Theorem~\ref{theorem_4} that the noise corruption in the input signal carries over to the modulo samples. % at which they have a closed-form relationship with each other. 
The following theorem unveils the UNO reconstruction guarantee in the presence of noise.

%-----------------------------------------------------------------
\begin{figure*}[t]
	\centering
	\includegraphics[width=1.0\columnwidth]{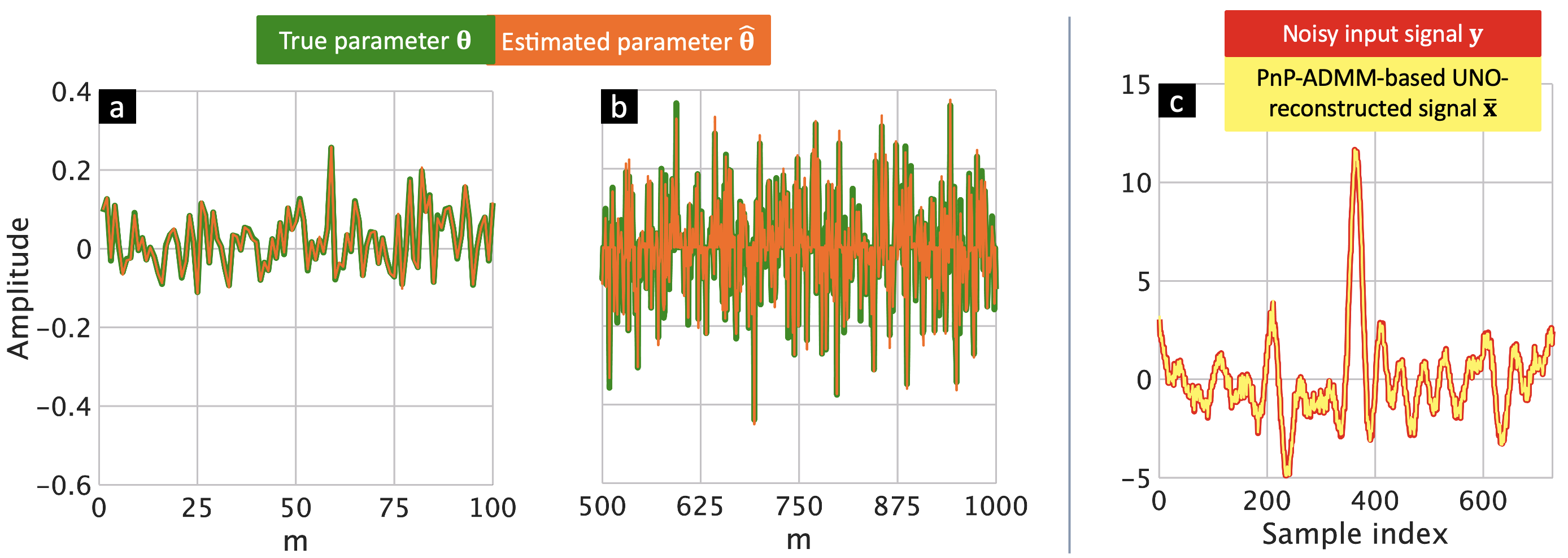}
	%\subfloat[Overdetermined System ($r\geq s$)]
	%	{\includegraphics[width=0.45\textwidth]{noisy-lasso-overdetermined.eps}}
	%\subfloat[Underdetermined system ($r\leq s$)]
	%	{\includegraphics[width=0.45\textwidth]{noisy-lasso-underdetermined.eps}}
	\caption{Reconstruction of the desired parameter vector $\btheta$ following the linear model (\ref{eq:1000}) using PnP-ADMM-based UNO for an %, where the true values are plotted alongside the estimates $\widehat{\btheta}$. 
	(a) overdetermined system with $\mbA\in\mathbb{R}^{728\times 100}$ and (b) underdetermined system with $\mbA\in\mathbb{R}^{728\times 1000}$. Here, to facilitate a better visual presentation, the number of threshold sequences start from $m=500$. (c) Reconstruction of the noisy input signal from one-bit measurements using PnP-ADMM-based UNO. 
    %\vspace{-10pt}
    }
	\label{figure_7}
\end{figure*}
%-----------------------------------------------------------------

\begin{theorem}
\label{theorem_5}(UNO sampling with noise)
Assume $x(t)$ to be a finite energy, bandlimited signal with maximum frequency $\Omega_{\textrm{max}}$. Let $y(t)$ denote the noisy signal following a linear model $y(t)=x(t)+z(t)$, where $z(t)$ denotes the additive noise signal. Denote the pre-filtered $y(t)$, $x(t)$ and $z(t)$ by, respectively, $y_{\phi}(t)$, $x_{\phi}(t)$ and $z_{\phi}(t)$, where $\phi\in\textrm{PW}_{\Omega}$ with cut-off frequency $\Omega_{\textrm{max}}$ following a linear model $y_{\phi}(t)=x_{\phi}(t)+z_{\phi}(t)$. Denote the samples of $y_{\phi}(t)$, $x_{\phi}(t)$, $z_{\phi}(t)$ and the modulo samples of $y_{\phi}(t)$ by, respectively, $\left(y_{\phi}\right)_{k}$, $\left(x_{\phi}\right)_{k}$, $\left(z_{\phi}\right)_{k}$ and $\left(\Tilde{y}_{\phi}\right)_{k}$, where the sampling rate is $1/T$. The modulo samples reconstructed by the RKA are denoted by $\bar{\Tilde{\mby}}_{\phi}$ with the reconstruction error is $\mbe=\left(\bar{\Tilde{\mby}}_{\phi}-\Tilde{\mby}_{\phi}\right)$. Then, the sufficient condition to reconstruct bandlimited samples $x_{k}$ from UNO samples $\left\{\mathbf{r}^{(\ell)}=\operatorname{sgn}\left(\mathcal{M}_{\lambda}\left(\mathbf{y}_{\phi}\right)-\boldsymbol{\uptau}^{(\ell)}\right)\right\}_{\ell=1}^{m}$, where $\boldsymbol{\uptau}^{(\ell)}\sim \mathcal{N}\left(\mathbf{0}, \frac{\lambda^{2}}{9}\mathbf{I}\right)$, up to additive multiples of $2\lambda$ in the sense that $\left(\bar{\Tilde{y}}_{\phi}\right)_{k}=x_{k}+\left(\Tilde{z}_{\phi}\right)_{k}+e_{k}$ is
\begin{equation}
\label{eq:n_5}
T \leq \frac{1}{2^{h}\Omega_{\textrm{max}} e},
\end{equation}
where $\left(\Tilde{z}_{\phi}\right)_{k}$ is defined in Theorem~\ref{theorem_4} and $h\in\mathbb{N}$ is determined by
\begin{equation}
\label{eq:n_6}  
h\geq \frac{\log\left(\frac{2\beta_{x}}{\lambda}\right)}{\log\left(\frac{\lambda}{4\left\|\Tilde{\mbz}_{\phi}+\mbe\right\|_{\infty}}\right)},
\end{equation}
with
\begin{equation}
\label{eq:n_7}
\lambda\geq 4\zeta\left\|\Tilde{\mbz}_{\phi}+\mbe\right\|_{\infty}, \quad \zeta>1.
\end{equation}
\end{theorem}
\begin{IEEEproof}
The proof, \textit{ceteris paribus}, follows from repeating the proof of Theorem~\ref{Negar_Theorem} by replacing the RKA error $\mbe$ with %the additive noise to the modulo samples 
$\Tilde{\mbz}_{\phi}+\mbe$.
\end{IEEEproof}
% According to Theorem~\ref{theorem_4}, we can conclude that (i) the input additive noise $z_{k}$ leads to a bounded noise $\Tilde{z}_{k}$ that is added to the modulo samples $\Tilde{\mathbf{x}}$, and (ii) the Noisy-UNO algorithm can guarantee the reconstruction of the desired signal $\mathbf{x}$ from noisy measurements $\mathbf{y}$, when the ADC threshold $\lambda$ obeys (\ref{th_1}). VIJAY: said this so many times. Just before the last para of the proof says so as well.
According to Theorem~\ref{theorem_5}, noisy UNO sampling requires more samples (given by the oversampling factor) specified by \eqref{eq:n_5} than the noiseless case in \eqref{rageN}. This is similar to other conventional noisy samplers. For example, Cadzow denoising \cite{cadzow1988signal}, used to suppress the effect of noise in sparse samplers similarly requires such an oversampling \cite{rudresh2017finite}.

\subsection{Numerical Examples}
We investigated PnP-ADMM-based noisy UNO reconstruction with $\mathbf{A}=\left[a_{ij}\right]$ to be $a_{ij}\sim\mathcal{N}\left(0,1\right)$ and %$\mathbf{y}$ follows 
$\mathbf{y}=\mathbf{y}_{t}+\bepsilon$, where $\mathbf{y}_{t}$ was generated as in Section~\ref{Guaranteed} and $\bepsilon\sim\mathcal{N}\left(\mathbf{0},\sigma^{2}_{\bepsilon} \mathbf{I}_{m}\right)$. Fig.~\ref{figure_7}a and b show accurate noisy UNO reconstruction of the parameter vector %with both desired parameter vector $\btheta$ depicted alongside its estimate $\widehat{\btheta}$ 
with fixed $\sigma^{2}_{\bepsilon}=0.1$ in case of, respectively, overdetermined ($r=728$, $s=100$) and underdetermined ($r=728$, $s=1000$) systems in (\ref{eq:1000}). %Fig.~\ref{figure_7} confirms the possibility of reconstructing $\btheta$ from noisy observations $\mathbf{y}$. 
Fig.~\ref{figure_7}c demonstrates the efficacy of Noisy UNO in estimating the desired parameter $\boldsymbol{\uptheta}$ from (\ref{eq:1000}) when only UNO samples of noisy measurement $\mby$ are available.

Table~\ref{table_4} reports the reconstruction NMSE of $\theta$, i.e.,
%\begin{equation}
%\label{nega17}
$\operatorname{NMSE}_{\boldsymbol{\uptheta}} \triangleq \frac{\left\|\boldsymbol{\uptheta}-\widehat{\boldsymbol{\uptheta}}\right\|_{2}^{2}}{\left\|\boldsymbol{\uptheta}\right\|_{2}^{2}}$,
%\end{equation}
averaged over $15$ experiments for different noise variances $\sigma^{2}_{\bepsilon}\in \left\{0.01,0.05,0.1\right\}$ using the PnP-ADMM-based UNO. Here, following Theorem~\ref{theorem_5}, the ADC threshold was set to $\lambda=1.5$. %We present the average of NMSE (\ref{eq:17}) over $15$ experiments for different noise variances.

%-----------------------------------------------------------------
%\begin{figure}[t]
%	\center{\includegraphics[width=0.45\textwidth]{compare_noisy_input.eps}}
%	\caption{Reconstruction of the noisy input signal from one-bit measurements using PnP-ADMM-based UNO.}
%	\label{figure_6}
%\end{figure}
%-----------------------------------------------------------------
%------------------------------------------
%\begin{table}[t]
%\centering
%\caption{Reconstruction of the desired parameter vector $\btheta$ by Noisy-UNO equipped with PnP-ADMM.}
%\centering
%\begin{tabular}{  c | c | c | c }
%\hline
%\textbf{Overdetermined system}&\textbf{$\sigma^{2}_{\bepsilon}=0.1$}& \textbf {$\sigma^{2}_{\bepsilon}=0.05$}&\textbf{$\sigma^{2}_{\bepsilon}=0.01$}\\ [0.5 ex]
%\hline \hline
%NMSE ($\log_{10}$ scale) & -4.8294 & -5.2676 & -5.6815\\
%\hline
%\textbf{Underdetermined system}&\text{$\sigma^{2}_{\bepsilon}=0.1$}& \text{$\sigma^{2}_{\bepsilon}=0.05$}&\text{$\sigma^{2}_{\bepsilon}=0.01$}\\
%\hline 
%NMSE ($\log_{10}$ scale) & -3.8128 & -4.2347 & -4.5259\\
%\hline
%\end{tabular}
%\label{table_4}
%\end{table}
%------------------------------------------
%-----------------------------------------------------------------
\begin{table} [t]
\centering
\caption{Reconstruction $10\log_{10}\operatorname{NMSE}_{\boldsymbol{\uptheta}}$ %of the desired parameter vector  $\btheta$ 
with PnP-ADMM Noisy UNO}
\centering
\begin{tabular}{ c | c | c }
\hline
$\sigma^{2}_{\bepsilon}$ & \textbf{Overdetermined system} & \textbf{Underdetermined system}\\ [0.5 ex]
\hline \hline
$0.1$ & $-48.294$ & $-38.128$ \\
\hline
$0.05$ & $-52.676$ & $-42.347$ \\
\hline
$0.01$ & $-56.815$ & $-45.259$ \\
\hline
\end{tabular}
%\vspace{-10pt}
\label{table_4}
\end{table}
%-----------------------------------------------------------------

\section{Summary}
\label{sec:summ}
The design of alternative sampling schemes to enable practical implementations of Shannon's theorem -- from theory to praxis -- has been an active research topic for decades. In this context, our proposed UNO presents a framework of merging one-bit quantization and unlimited sampling. This sampling framework naturally facilitates a judicious design of time-varying sampling thresholds by properly utilizing the information on the distance between the signal values and the thresholds in a high DR regime. The noiseless UNO reconstruction relies on exploiting RKA algorithm while the noisy reconstruction is based on the PnP-ADMM heuristic. These low-complexity approaches are preferable over existing costly reconstruction optimization approaches \cite{venkatakrishnan2013plug,chan2016plug}. 

The UNO framework achieves multiple objectives of high sampling rate, unlimited DR, less complex and potentially low-power implementations. Our numerical and theoretical analyses demonstrate accurate reconstruction for several different scenarios. Some theoretical questions remain open, e.g. on the relationship between the number of threshold sequences $m$ and reconstruction error in a closed form. This may help in finding the required number of thresholds sequences for perfect reconstruction. Further, a hardware verification of UNO on the lines of unlimited sampling in \cite{bhandari2021unlimited} is also desired.
%\vspace{-12pt}
\section*{Acknowledgement}
The authors are grateful to Prof. Ayush Bhandari of Imperial College, London for helpful discussions related to his work in \cite{graf2019one} and for providing the relevant source codes to facilitate meaningful comparison.

%\newpage
\bibliographystyle{IEEEtran}
\bibliography{references_updated_from_ICASSP}

\begin{thebibliography}{10}
\bibitem{whittaker1915functions}
E.~T. Whittaker, ``On the functions which are represented by the expansion of
  interpolating theory,'' \emph{Proceedings of the Royal Society}, vol.~35, pp.
  181--194, 1915.
\end{thebibliography}

\end{document}